\documentclass{aastex63}
\usepackage{amsmath}
\usepackage{mathdots}
\usepackage{natbib}
\usepackage[utf8]{inputenc}
\usepackage{lineno}
\usepackage{chngcntr}
\definecolor{darkblue}{RGB}{0,0,196}
\definecolor{darkgreen}{RGB}{0,120,0}
\received{Jan 14, 2021}
\accepted{\today}
\submitjournal{\apj}

\shorttitle{Observational study of solar surface convection}
\shortauthors{Kashyap and Hanasoge}
\graphicspath{{./}{figures/}}
\renewcommand\sout{\bgroup\markoverwith{\textcolor{blue}{\rule[0.5ex]{2pt}
      {0.4pt}}}\ULon}

\renewcommand{\vec}[1]{\boldsymbol{#1}}

\newcommand{\vecr}{\vec{r}}
\newcommand{\los}{\hat{\vec{\ell}}}

\begin{document}
\title{Characterising solar surface convection using Doppler measurements}
\correspondingauthor{Samarth G. Kashyap}
\email{g.samarth@tifr.res.in}

\author[0000-0001-5443-5729]{Samarth G. Kashyap}
\affil{Department of Astronomy and Astrophysics \\
Tata Institute of Fundamental Research \\
Mumbai, India}

\author[0000-0003-2896-1471]{Shravan M. Hanasoge}
\affil{Department of Astronomy and Astrophysics \\
Tata Institute of Fundamental Research \\
Mumbai, India}

\begin{abstract}
The Helioseismic Magnetic Imager (HMI) onboard the 
\emph{Solar Dynamics Observatory} (SDO) records line-of-sight Dopplergram images of convective flows on the surface. These images are used to obtain 
the multi-scale convective spectrum. We design a pipeline to process the raw images 
to remove large-scale features like differential rotation, 
meridional circulation, limb shift and imaging artefacts. 
The Hierarchical Equal Area Pixelization scheme (HEALPix)
is used to perform spherical harmonic transforms on the cleaned image.
{
Because we only have access to line-of-sight velocities on 
half the solar surface, we define a ``mixing matrix" to relate 
the observed and true spectra. 
This enables the inference of poloidal and toroidal flow spectra 
in a single step through the inversion of the mixing matrix.
Performing inversions on a number of flow profiles, we find that
the poloidal flow recovery is most reliable among all the components.
We also find that the poloidal spectrum is in qualitative agreement 
with inferences from Local Correlation Tracking (LCT) of granules.
The fraction of power in vertical motions increases as a function
of wavenumber and is at 8\% level for $\ell = 1500$. In contrast to seismic results and LCT, the flows show nearly no  temporal-frequency dependence.  Poloidal flow power peaks in the range of 
$\ell - |m| \approx 150-250$ which may potentially hint at a latitudinal preference for convective
flows.
}
\end{abstract}

%------------------------begin sec:intro--------------------------------
\section{Introduction}\label{sec:intro}
The Sun, like many stars in the main-sequence,  possesses a convectively
unstable outer envelope surrounding an inner stable core. 
Given its proximity to us, we observe the Sun with high spatial and 
temporal resolution. 
Measurements of plasma flows on the surface reveal a highly intricate 
and dynamic cellular structure, highlighting a range of spatio-temporal 
scales \citep[e.g.][]{miesch2005lsrp,rast2003}. 
Solar convection, which is primarily driven by buoyancy and radiative cooling
at the surface, manifests in an overturning of fluid between the base of the
convection zone and the solar surface.
Since the fluid is ionized everywhere
\citep[almost fully ionized for 97\% solar radius and 
partially ionized in the top 3\%; ][]{nordlund2009lsrp}, it serves to carry magnetic field, playing
a crucial role in setting solar dynamics. Hence,
an understanding of solar convection will enable us to better appreciate thermal and angular momentum transport \citep{aerts2019},
ultimately bringing us closer to explaining the 11-year solar magnetic cycle,
a long-standing problem in solar physics.

Solar convection plays out in an extreme parameter regime; the Reynolds number, $Re \sim 10^{14}$,
and the Prandtl number, $Pr \sim 10^{-7}$. 
These regimes are inaccessible to experiments and hence, 
attempts to understand the physics of turbulent convection of the 
Sun involve numerical analysis. However, conservative estimates of 
the viscous dissipation scale of the Sun place it at $\sim 100$ m \citep{brummell95,lesieur-metais-96},
six orders of magnitude smaller than the largest length scale i.e the radius of the Sun, making direct numerical simulation 
\citep{orszag1970,moin-mahesh-1998} intractable. Hence the inclination to use
different types of numerical simulations to study different phenomenon: 
(a) high-resolution local simulations are used to study dynamics of near-surface layers 
\citep{stein-nordlund-1998, rincon2005} and (b) large-scale flows such as differential rotation
are studied using global models based on the anelastic equations
\citep{cattaneo2003, derosa-gilman-toomre-02} etc.
Simulations attempt to reproduce different aspects of observed solar convection.
Hence robust inferences from observational data become a necessity, not only as 
a standalone measurement by itself, but also as a reference point for validation of
numerical simulations. 

Observational studies of convection have found three prominent spatial
scales. Granulation is a result of vertical advection and downward plume formation 
driven by radiative losses at the photosphere. 
Granules are found to occur at spatial scales of $\sim 1$ Mm, with 
typical lifetimes of $0.2$ hr \citep{herschel-1801, richardson-schwarzschild-1950, leighton62}.
Although mesogranulation was initially 
observed at a length scale of $\sim 5$ Mm \citep{november-etal-81,muller92}
and a lifetime of $\sim 3$ hours, full-disk observations
from HMI have established the absence of peak in 
convective power at those scales \citep{hathaway2015, 
rincon-2017}.
%and is thought to be due to ``collective interaction 
%of granular plumes'' \citep{rast2003}.
Supergranulation is observed at $\sim 30$ Mm, which advects granular structures \citep{muller92,rieutord2010}.

%Modelling solar convection in spherical geometry --surface
%shear layer established by helioseismology. Simulations used to suggest mechanism to
%maintain radial shear.

%  \item \tr{Limitations of other techniques}
%\cite{duvall1980} showed that supergranulation cells rotate faster than local
%plasma by nearly 4\%.
\cite{rieutord2001} showed that granules serve as good tracers of local plasma
flows at meso- and supergranular scales over long time scales. Local correlation
tracking of granules \citep{november_simon1988} was used to
study convection by \cite{georgobiani2006} in granular to supergranular-scale
simulations and they found the velocity spectrum to be approximately a power
law for scales larger than granules. \cite{rieutord2010} studied the high-wavenumber power spectrum using granulation tracking for studying horizontal
and vertical velocity by measuring Doppler velocity at disk
center. Estimates of the convection spectrum of
radial, toroidal and poloidal flows were made by \cite{hathaway2015} (H15 hereafter) by employing a different method for separately
estimating these flows. They noted that the spectrum was dominated by 
poloidal flows for spherical harmonic degree $\ell > 30$ and toroidal flows for 
$\ell < 30$.

The aim of the work is to estimate the vector velocity spectrum on the
surface of the Sun, using line-of-sight Dopplergrams. HMI provides us with Dopplergrams measured with a \(720\)s cadence. 
Since we measure only the line-of-sight component of 
velocity on less than half the surface of the
Sun, there is a loss of information. Hence, projecting the observed
velocities on to any orthonormal basis on the surface of the Sun is imperfect.
This is  quantified using a mixing-matrix which relates the ``true'' spectral 
parameters to their observed counterparts. 
Previously, \cite{rincon-2017} obtained robust vector velocity 
spectrum using coherent structure tracking analysis (CST) of granules
\citep{roudier-2012}. The uniqueness
in our approach is the ability to obtain the toroidal and
poloidal (both vertical and horizontal) convection spectrum 
from the line-of-sight Dopplergrams directly. This enables us to 
probe the spectral features of the tiniest scales of the observation.
The 16 megapixel Dopplergram images from the HMI enable us to obtain 
spectra upto $\ell$ = 1545.
Another novelty of this work is the usage of HEALPix
\footnote{HEALPix website: \url{http://healpix.sf.net/}} \citep{healpix} for constructing pixels on the solar 
surface, instead of the traditional Gaussian collocated grid.
The Dopplergram images have more pixels 
near the equator than the poles. 
Irrespective of longitude, the traditionally used Gaussian 
collocated grid has the same number of pixels along longitude near the 
poles and the equator. This results in super-sampling near
the poles and under-sampling near the equator. Thus, HEALPix
provides us with a natural pixelization on the sphere for
solar Dopplergram data.

%\tr{We argue that it is better suited for analyzing datasets from HMI.}
%  \item \tr{Structure of our paper}

The outline of the paper is as follows. The theoretical background about spectral
decomposition and spectral mixing are dealt with in Section~\ref{sec:theory}. Before 
applying the present method to HMI data, the raw data need to be cleaned to remove
systematical errors, described in Section~\ref{sec:data-prep}. We
discuss corrections due to the motion of HMI in Section~\ref{subsec:obs-motion}, compensating
for gravitational redshift in Section~\ref{subsec:grav-redshift} and removal of large-scale
features in Section~\ref{subsec:conv-bs}. The analysis of cleaned HMI data is 
discussed in Section~\ref{sec:data-analysis}. The validation of the 
inversion procedure is discussed in Section~\ref{sec:goodness-inversion}.
The results are discussed in Section~\ref{sec:results-discussion}.

\section{Theoretical Background}\label{sec:theory}
The Doppler velocity on the surface \(\tilde{\vec{u}}(\theta, \phi)\)
may be expanded in the vector spherical harmonic (VSH) basis  \( (\vec{Y}_{\ell m}
, \vec{\Psi}_{\ell m}, \vec{\Phi}_{\ell m}) \) according as
\begin{equation}\label{eq:vecsph}
  \tilde{\vec{u}}(\theta, \phi)
  = \sum_{\ell m} \left( \tilde{u}_{\ell m} \vec{Y}_{\ell m}(\theta, \phi)
            + \tilde{v}_{\ell m} \vec{\Psi}_{\ell m}(\theta, \phi)
            + \tilde{w}_{\ell m} \vec{\Phi}_{\ell m}(\theta, \phi) \right).
\end{equation}
Definitions and properties of the basis are given in Appendix
\ref{sec:vsh}. We use both  $(\ell, m)$ and $(s, t)$ to denote 
spherical harmonic degree and the azimuthal order respectively.
The $w_{\ell m}$ component is called the toroidal component and 
$(u_{\ell m}, v_{\ell m})$ are together called as the poloidal 
component. For the sake of differentiating, we would use the term
``radial'' to refer to the poloidal-vertical component $u_{\ell m}$ 
and ``poloidal'' to refer to poloidal-horizontal component $v_{\ell m}$.
If we observe all components of velocity on the entire solar surface, then
we can perfectly isolate all VSH components
\(\tilde{U}_{\ell m} = (\tilde{u}_{\ell m}, \tilde{v}_{\ell m}, \tilde{w}_{\ell m})\), by
exploiting the orthonormality of the VSH basis. 
However, we observe only the velocity projected on to the line-of-sight vector
\(\los(\theta, \phi)\), 
\begin{equation}
  \vec{u}(\theta, \phi) = \left[\los(\theta, \phi) \cdot \tilde{\vec{u}}(\theta, \phi)\right]
  \los(\theta, \phi),
\end{equation}
over less than half of the solar surface. We use tilde to denote
the true spectral components of velocity and to distinguish them from
the spectral components of the line-of-sight projected velocity.
Hence, observed velocities in spectral domain, \(U_{st} = (u_{st}, 
v_{st}, w_{st})\) 
are given by Eqn.~(\ref{eqn:ulm}, \ref{eqn:vlm}, \ref{eqn:wlm}),
\begin{align}
 u_{st} & = \int d\Omega \, W(\theta, \phi)\,  \vec{u} \cdot \vec{Y}^*_{st} \label{eqn:ulm}\\
 v_{st} & = \int d\Omega \, W(\theta, \phi)\,  \vec{u} \cdot \vec{\Psi}^*_{st} \label{eqn:vlm}\\
 w_{st} & = \int d\Omega \, W(\theta, \phi)\,  \vec{u} \cdot \vec{\Phi}^*_{st} \label{eqn:wlm},
\end{align}
where \(W(\theta, \phi)\) is a window function which has value \(1\)
where the Sun is observed and \(0\) otherwise.
Using Eqn.~(\ref{eq:vecsph}), we write the observed
spectral components $U_{st}$ in terms of the true spectral components of
velocity $\tilde{U}_{\ell m}$,
\begin{equation}\label{eqn:leak1comp}
 u_{st} = \sum_{\ell} \sum_{m=-\ell}^{\ell}\left( 
 \tilde{u}_{\ell m} M^{(uu)\ell m}_{st} +
 \tilde{v}_{\ell m} M^{(uv)\ell m}_{st} + 
 \tilde{w}_{\ell m} M^{(uw)\ell m}_{st}\right),
\end{equation}
\begin{equation}\label{eqn:leak2comp}
 v_{st} = \sum_{\ell} \sum_{m=-\ell}^{\ell} \left( 
 \tilde{u}_{\ell m} M^{(vu)\ell m}_{st} +
 \tilde{v}_{\ell m} M^{(vv)\ell m}_{st} + 
 \tilde{w}_{\ell m} M^{(vw)\ell m}_{st} \right),
\end{equation}
\begin{equation}\label{eqn:leak3comp}
 w_{st} = \sum_{\ell} \sum_{m=-\ell}^{\ell} \left( 
 \tilde{u}_{\ell m} M^{(wu)\ell m}_{st} +
 \tilde{v}_{\ell m} M^{(wv)\ell m}_{st} + 
 \tilde{w}_{\ell m} M^{(ww)\ell m}_{st} \right).
\end{equation}
In Eqns.~(\ref{eqn:leak1comp}, \ref{eqn:leak2comp}, \ref{eqn:leak3comp}), the matrix $M_{\ell m}^{st}$ 
quantifies the extent of mixing of true spectral coefficients
corresponding to quantum numbers given by $(\ell, m)$ into 
observed spectral coefficients specified by $(s, t)$. 
Note that the mixing occurs not only between quantum
numbers, but also among the radial, poloidal and 
toroidal (VSH) components. 
This is indicated by the superscript parenthesis. 
For instance, the mixing matrix $M^{(vw)\ell m}_{st}$
quantifies the contribution of the true toroidal 
component ($w_{st}$) towards the observed poloidal 
component ($v_{\ell m}$). The mixing matrices
are given in Eqn.~(\ref{eqn:mixmat-compact}),
\begin{equation}
 \label{eqn:mixmat-compact}
 M^{(ij)\ell m}_{st} = \int d\Omega \, W(\theta, \phi)\,
 \mathcal{J}^{(ij)}_{\ell m} \mathcal{K}^{(ij)*}_{st},
\end{equation}
where $i, j$ are variables which can be any of $u, v, w$. The set of all $\mathcal{J}, \mathcal{K}$
are given in Table~(\ref{tab:mixlist})
\begin{table}[!h]
\label{tab:mixlist}
\caption{Table of $\mathcal{J}$ and $\mathcal{K}$}
\centering
\begin{tabular}{|c|c|c|c|}
\hline
  & $i=u$ & $i=v$ & $i=w$ \\
\hline
$j=u$ & 
$\mathcal{J}^{(ij)}_{\ell m} = \los \cdot \vec{Y}_{\ell m}$; 
$\mathcal{K}^{(ij)*}_{st} = \los \cdot \vec{Y}^*_{st}$
& 
$\mathcal{J}^{(ij)}_{\ell m} = \los \cdot \vec{Y}_{\ell m}$; 
$\mathcal{K}^{(ij)*}_{st} = \los \cdot \vec{\Psi}^*_{st}$
& 
$\mathcal{J}^{(ij)}_{\ell m} = \los \cdot \vec{Y}_{\ell m}$; 
$\mathcal{K}^{(ij)*}_{st} = \los \cdot \vec{\Phi}^*_{st}$
\\
\hline
$j=v$ & 
$\mathcal{J}^{(ij)}_{\ell m} = \los \cdot \vec{\Psi}_{\ell m}$; 
$\mathcal{K}^{(ij)*}_{st} = \los \cdot \vec{Y}^*_{st}$
& 
$\mathcal{J}^{(ij)}_{\ell m} = \los \cdot \vec{\Psi}_{\ell m}$; 
$\mathcal{K}^{(ij)*}_{st} = \los \cdot \vec{\Psi}^*_{st}$
& 
$\mathcal{J}^{(ij)}_{\ell m} = \los \cdot \vec{\Psi}_{\ell m}$; 
$\mathcal{K}^{(ij)*}_{st} = \los \cdot \vec{\Phi}^*_{st}$
\\
\hline
$j=w$ & 
$\mathcal{J}^{(ij)}_{\ell m} = \los \cdot \vec{\Phi}_{\ell m}$; 
$\mathcal{K}^{(ij)*}_{st} = \los \cdot \vec{Y}^*_{st}$
& 
$\mathcal{J}^{(ij)}_{\ell m} = \los \cdot \vec{\Phi}_{\ell m}$; 
$\mathcal{K}^*_{\ell m} = \los \cdot \vec{\Psi}^*_{st}$
& 
$\mathcal{J}^{(ij)}_{\ell m} = \los \cdot \vec{\Phi}_{\ell m}$; 
$\mathcal{K}^{(ij)*}_{st} = \los \cdot \vec{\Phi}^*_{st}$
\\
\hline
\end{tabular}
\end{table}
%
%\begin{align}
% M^{(uu)st}_{\ell m} & = \int d\Omega \, W(\theta, \phi) \, \left(\los \cdot \vec{Y}_{st}\right) 
%\left( \los \cdot \vec{Y}^*_{\ell m} \right) \label{eqn:muu} \\
% M^{(uv)st}_{\ell m} & = \int d\Omega \, W(\theta, \phi) \, \left(\los \cdot \vec{\Psi}_{st}\right) 
%\left( \los \cdot \vec{Y}^*_{\ell m} \right) \label{eqn:muv} \\
% M^{(uw)st}_{\ell m} & = \int d\Omega \, W(\theta, \phi) \, \left(\los \cdot \vec{\Phi}_{st}\right) 
%\left( \los \cdot \vec{Y}^*_{\ell m} \right) \label{eqn:muw} \\
%
% M^{(vu)st}_{\ell m} & = \int d\Omega \, W(\theta, \phi) \, \left(\los \cdot \vec{Y}_{st}\right) 
%\left( \los \cdot \vec{\Psi}^*_{\ell m} \right) \label{eqn:mvu} \\
% M^{(vv)st}_{\ell m} & = \int d\Omega \, W(\theta, \phi) \, \left(\los \cdot \vec{\Psi}_{st}\right) 
%\left( \los \cdot \vec{\Psi}^*_{\ell m} \right) \label{eqn:mvv} \\
% M^{(vw)st}_{\ell m} & = \int d\Omega \, W(\theta, \phi) \, \left(\los \cdot \vec{\Phi}_{st}\right) 
%\left( \los \cdot \vec{\Psi}^*_{\ell m} \right) \label{eqn:mvw} \\
%
% M^{(wu)st}_{\ell m} & = \int d\Omega \, W(\theta, \phi) \, \left(\los \cdot \vec{Y}_{st}\right) 
%\left( \los \cdot \vec{\Phi}^*_{\ell m} \right) \label{eqn:mwu} \\
% M^{(wv)st}_{\ell m} & = \int d\Omega \, W(\theta, \phi) \, \left(\los \cdot \vec{\Psi}_{st}\right) 
%\left( \los \cdot \vec{\Phi}^*_{\ell m} \right) \label{eqn:mwv} \\
% M^{(ww)st}_{\ell m} & = \int d\Omega \, W(\theta, \phi) \, \left(\los \cdot \vec{\Phi}_{st}\right) 
%\left( \los \cdot \vec{\Phi}^*_{\ell m} \right) \label{eqn:mww}
%\end{align}
The linear relationships between the observed spectral components 
$U_{st}$ and the true spectral components $\tilde{U}_{\ell m}$ 
are given by Eqn.~(\ref{eqn:ulm} -- \ref{eqn:wlm}). 
These relationships may be combined into a single equation by 
defining vectors $\mathcal{U}$, $\tilde{\mathcal{U}}$ in the 
following manner,
\begin{align}
 \mathcal{U} & =
 \begin{bmatrix}
  \, \, \, \hdots \, \, \, u_i \, \, v_i \, \, w_i \, \, \, \hdots \, \, \,
 \end{bmatrix}^T \label{eqn:vecu} \\
 \tilde{\mathcal{U}} & =
 \begin{bmatrix}
  \, \, \, \hdots \, \, \, \tilde{u}_i \, \, \tilde{v}_i \, \, \tilde{w}_i \, \, \, \hdots \, \, \,
 \end{bmatrix}^T, \label{eqn:vecu-tilde}
\end{align}
where \(i\) is a combined index denoting the quantum numbers \((\ell, m)\). 
\begin{equation}
 \begin{bmatrix} \vdots\\ u_i \\ v_i \\ w_i \\ \vdots \end{bmatrix} =
 \begin{bmatrix}
  \ddots & \vdots & \vdots & \vdots & \iddots \\
  \dots  & M^{(uu)j}_i & M^{(uv)j}_i & M^{(uw)j}_i & \dots \\
  \dots  & M^{(vu)j}_i & M^{(vv)j}_i & M^{(vw)j}_i & \dots \\
  \dots  & M^{(wu)j}_i & M^{(wv)j}_i & M^{(ww)j}_i & \dots \\
  \iddots & \vdots & \vdots & \vdots & \ddots
 \end{bmatrix}
 \begin{bmatrix} \vdots\\ \tilde{u}_j \\ \tilde{v}_j \\ \tilde{w}_j \\ \vdots \end{bmatrix},
 \label{eqn:mixing-matrix}
\end{equation}
which may be written in compact form as
\begin{equation}
 \mathcal{U} = \mathcal{M} \tilde{\mathcal{U}}.
 \label{eqn:mixing-matrix-compact}
\end{equation}

\section{Data Preprocessing}\label{sec:data-prep}
\begin{figure}[!ht]
\centering
\includegraphics[width=\linewidth]{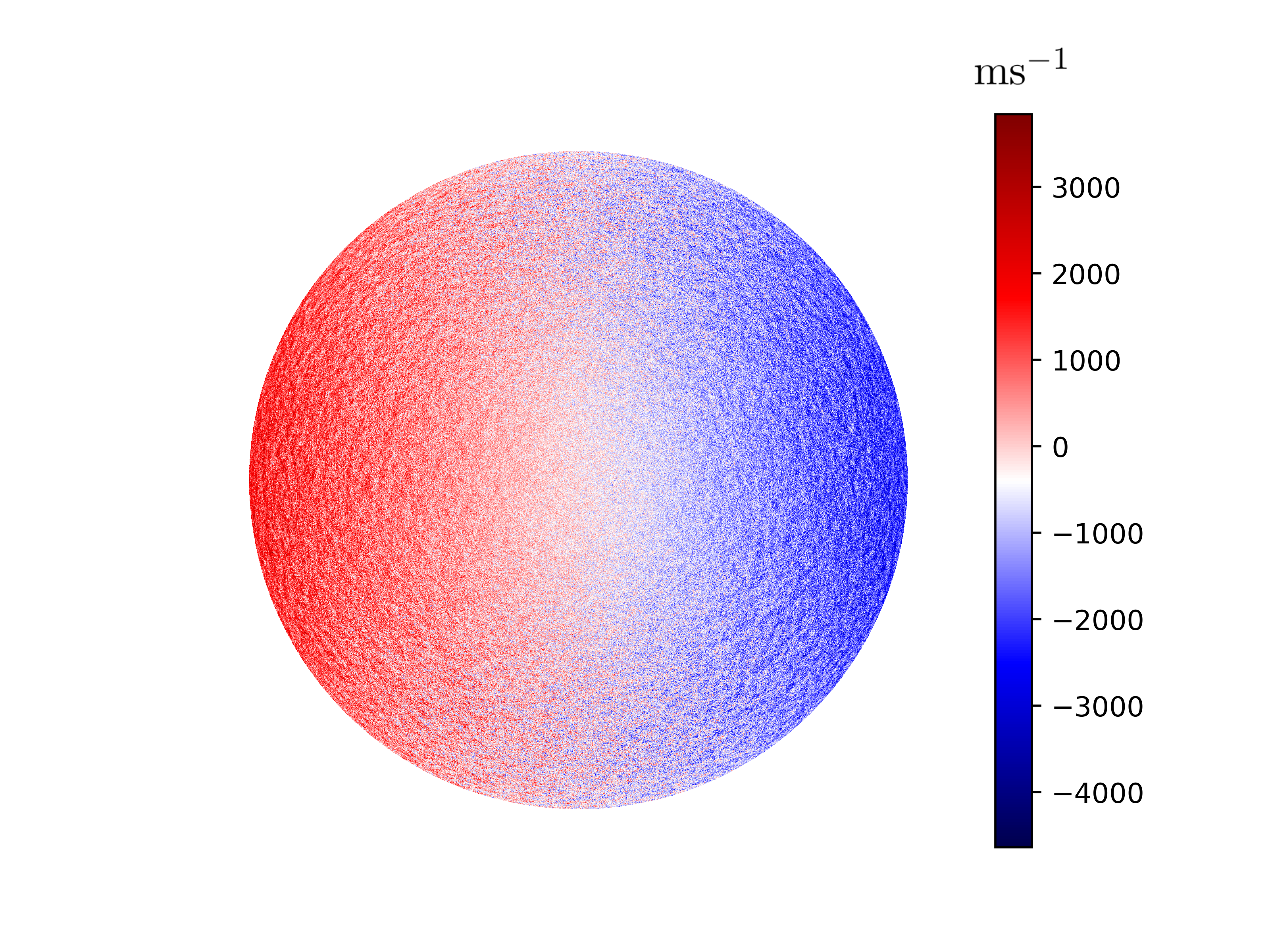}
\caption{Map of raw Dopplergram after the pixels outside $0.95 r_{disk}$ have been removed. The maximum
value is $+3840$ ms$^{-1}$ and the minimum value is $-4636$ ms$^{-1}$}
\label{fig:rawMap}
\end{figure}

For this study of surface convection, we use line-of-sight Dopplergrams
recorded by HMI, which observes the full solar disk at a resolution of 1 arc
second \citep{hmi2012}. Each Dopplergram is constructed using 72
filtergrams across the Fe-I line (6173.3 $\AA$) at a cadence of 720
seconds. The HMI data pipeline also provides us with a Dopplergram deconvolved
with the Point Spread Function (PSF) of the instrument, through the series
\textit{hmi.V\_720s\_dConS} (1 image per day), which is what we use for this
study. The raw image obtained from HMI is shown in Fig.~\ref{fig:rawMap}.

The preprocessing of the raw data involves removal of the following large-scale features
which are spurious to the present study:
\begin{enumerate}
 \item Motion of the observer (HMI).
 \item Gravitational redshift.
 \item Convective blueshift.
 \item Differential rotation.
 \item Meriodional circulation.
 \item Imaging artefacts.
\end{enumerate}
%As a part of this work, a pipeline was developed in Python to remove these 
%artefacts \citep{sgkashyap2020}.

% Motion of the observer
\subsection{Motion of the Observer}\label{subsec:obs-motion}
\begin{figure}[!ht]
\centering
\includegraphics[width=0.6\linewidth]{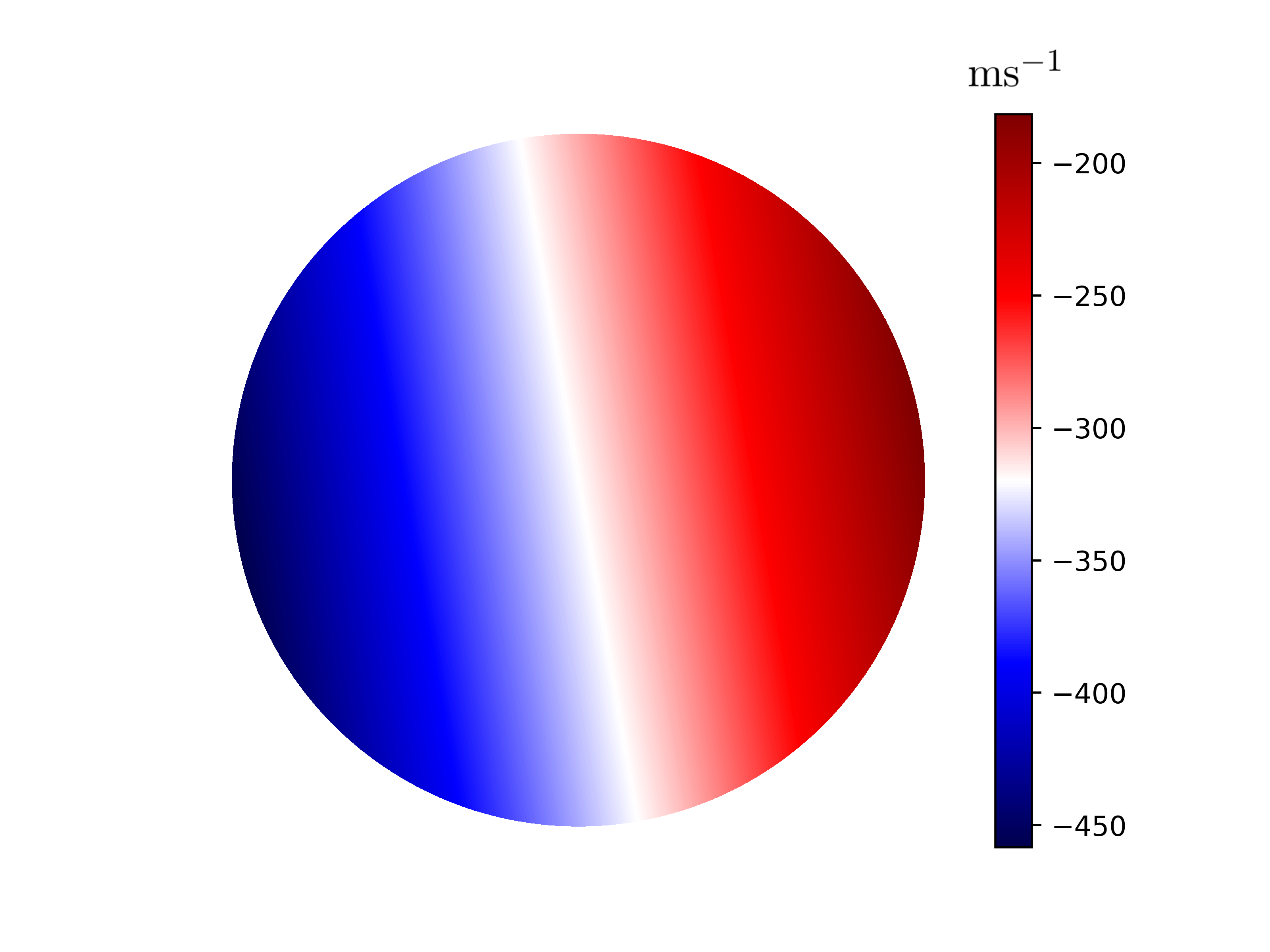}
\caption{Correction to the Dopplergram due to motion of
the observer, in ms$^{-1}$. The dominant effect is seen in the 
east-west direction, as the space-based SDO has the 
largest component of velocity in the direction of  its orbital motion.}
\label{fig:velCorr}
\end{figure}

The Dopplergram measures the relative velocity between the observer and the
source. To measure the velocity on the solar surface in the rest frame of 
the Sun, the velocity of the spacecraft needs to be accounted for.
The HMI data file provides the observer velocity through the keywords 
\verb|OBS_VR, OBS_VN, OBS_VW| for velocities in the radial, 
north-south and east-west directions respectively. In angular
heliocentric coordinates, \((\sigma, \chi)\), 
we write the velocity correction as Eqn.~(\ref{eqn:velCorr}),
\begin{equation}
 v_{obs} = V_R \cos(\sigma) - V_N \sin(\sigma) \cos(\chi) - V_W \sin(\sigma) \sin(\chi) \label{eqn:velCorr}.
\end{equation}
This correction is shown over the solar disk in Fig.~\ref{fig:velCorr}.

\subsection{Gravitational Redshift}\label{subsec:grav-redshift}
\begin{figure}[!ht]
\centering
\includegraphics[width=0.8\linewidth]{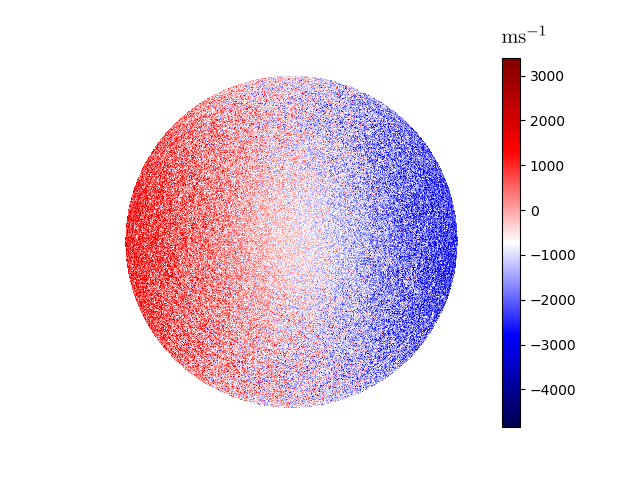}
\caption{Dopplergram after removal of effects due to motion of the
observer and gravitational redshift.}
\label{fig:afterVelCorr}
\end{figure}

Gravitational redshift is the consequence of the source
(solar surface) being at a higher gravitational potential than the observer
(HMI). To leading order, the density profile of the standard solar model \citep{jcd_models}
is spherically symmetric. It was subsequently shown by \cite{basu96b} 
that deviations of density from the solar model are at most $1.5$\%
and the deviations are also a function of only the radius, implying that the gravitational 
potential is also only a function of radius. Therefore, the 
surface of the Sun is at the same gravitational potential irrespective 
of the angular coordinates $(\theta, \phi)$. 
The redshift computed using Einstein's principle of equivalence 
for an observer at infinity is found to be $636$ ms$^{-1}$, which
is in agreement with observations \citep{beckers77,lopresto91,cacciani06}.
The Dopplergram, after removal of contributions due to observer velocity
and gravitational redshift, is shown in Figure~\ref{fig:afterVelCorr}.
%\footnote{$636$ ms$^{-1}$ for an observer at infinity 
%\tb{https://arxiv.org/pdf/1207.0177.pdf}}}

\subsection{Convective Blueshift, Differential Rotation and Meridional Circulation}
\label{subsec:conv-bs}
Outside of localized strong magnetic field concentrations, granules 
dot the entirety of the solar surface. Hot plasma outflows from the 
cell centers radiate heat and cold plasma plunges into the solar 
interior through the much narrower intergranular
lanes. Hence, at these scales, we would observe a blueshift in the bulk of the
granule and redshift at the intergranular lanes. The granules also appear
brighter owing to their higher temperatures. Hence, unresolved granules
throughout the surface of the solar disk gives rise to an apparent blue shift,
which is known as convective blueshift or limb-shift 
\citep{beckers1978,dravins1982}.
The limb-shift is represented by a polynomial of degree $5$ in heliocentric
coordinates \citep{thompson2006}.

\begin{equation}
 \los \cdot \vec{u}_{\text{LS}} = \sum_{i=0}^5 c_i P_i(x),
\end{equation}
where $\vec{u}_{\text{LS}}$ is the limb-shift, $\los$ is the line-of-sight
vector, \(x = 1 - \cos\rho\), \(\rho\) being the heliocentric angle and
\(P_i\) are the shifted Legendre polynomials (Appendix \ref{sec:legpoly}). 

\begin{figure}[!ht]
\centering
\includegraphics[width=\linewidth]{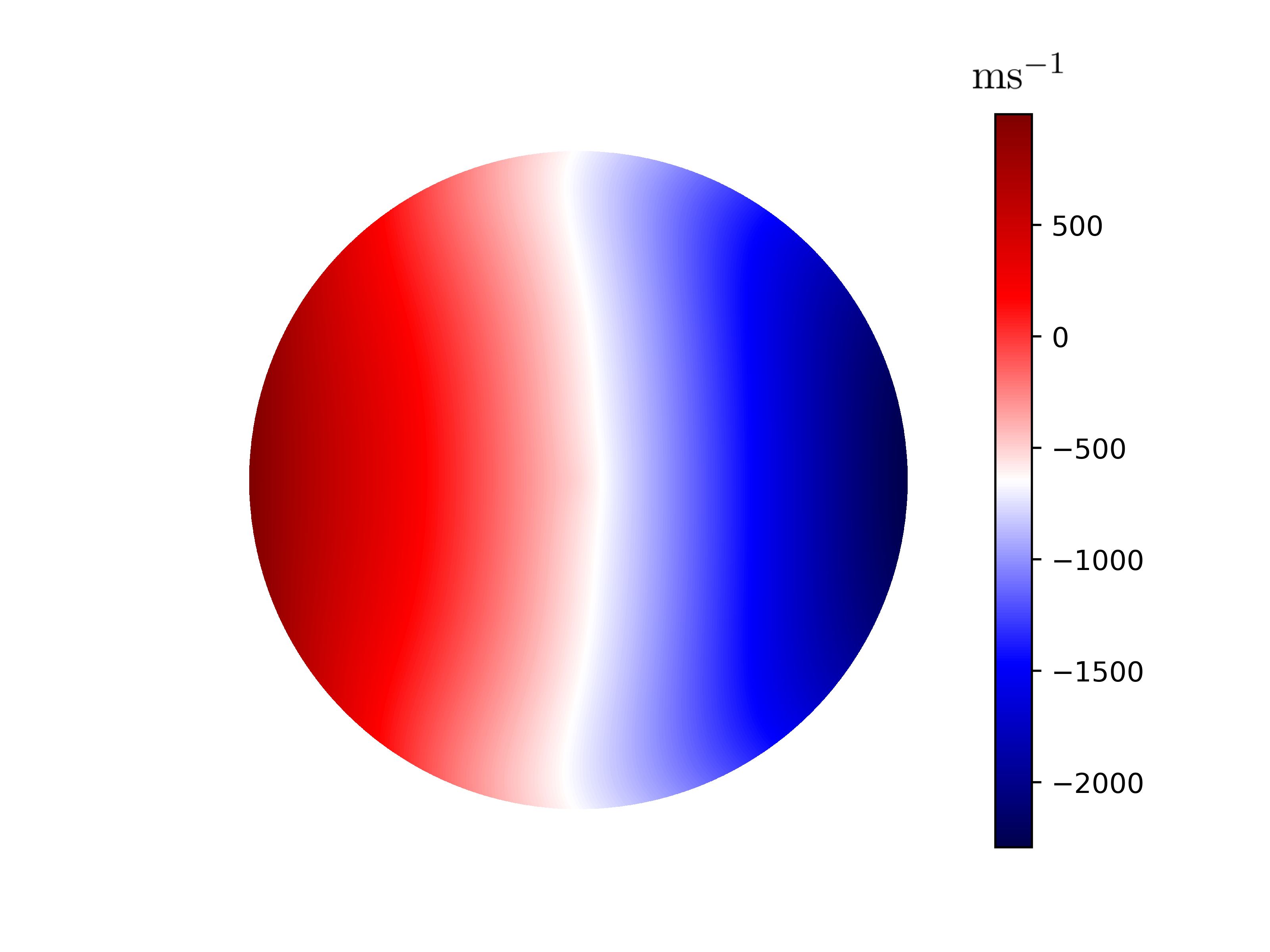}
\caption{Map showing the combined effect of differential rotation, meridional
circulation and convective blueshift. The map has a greater amount of blueshift
(\(\approx -2500 \) ms\(^{-1}\)) than redshift (\(\approx 550\) ms\(^{-1}\))
because of the limb-shift correction to be a blueshift all over the solar disk
whereas the correction due to differential rotation is red and blue shifted by
equal magnitudes.}
\label{fig:fitted}
\end{figure}

Differential rotation is an axisymmetric toroidal component of 
flow, which is symmetric about the equator. Meridional circulation 
is an axisymmetric poloidal flow which is antisymmetric 
about the equator. Hence differential rotation
is expressed in terms of the odd-degree toroidal 
components and meridional circulation using even-degree poloidal components, given by
Eqn.~(\ref{eqn:rot}, \ref{eqn:MC}),

\begin{equation}
 \vec{u}_{\text{DR}} = \sum_{\ell=1, 3, 5} w_{\ell 0}\, \hat{\vec{r}} \times \vec{\nabla}_h Y_{\ell 0},
 \label{eqn:rot}
\end{equation}

\begin{equation}
 \vec{u}_{\text{MC}} = \sum_{\ell=2, 4} v_{\ell 0}\, 
 \vec{\nabla}_h Y_{\ell 0}
 \label{eqn:MC}.
\end{equation}
We perform a least-squares fit to estimate the values of 
\(w_{\ell 0}\) and \( c_i\).
The total sum of all three components is shown in Figure~\ref{fig:fitted}.

\begin{figure}[!ht]
\centering
\includegraphics[width=\linewidth]{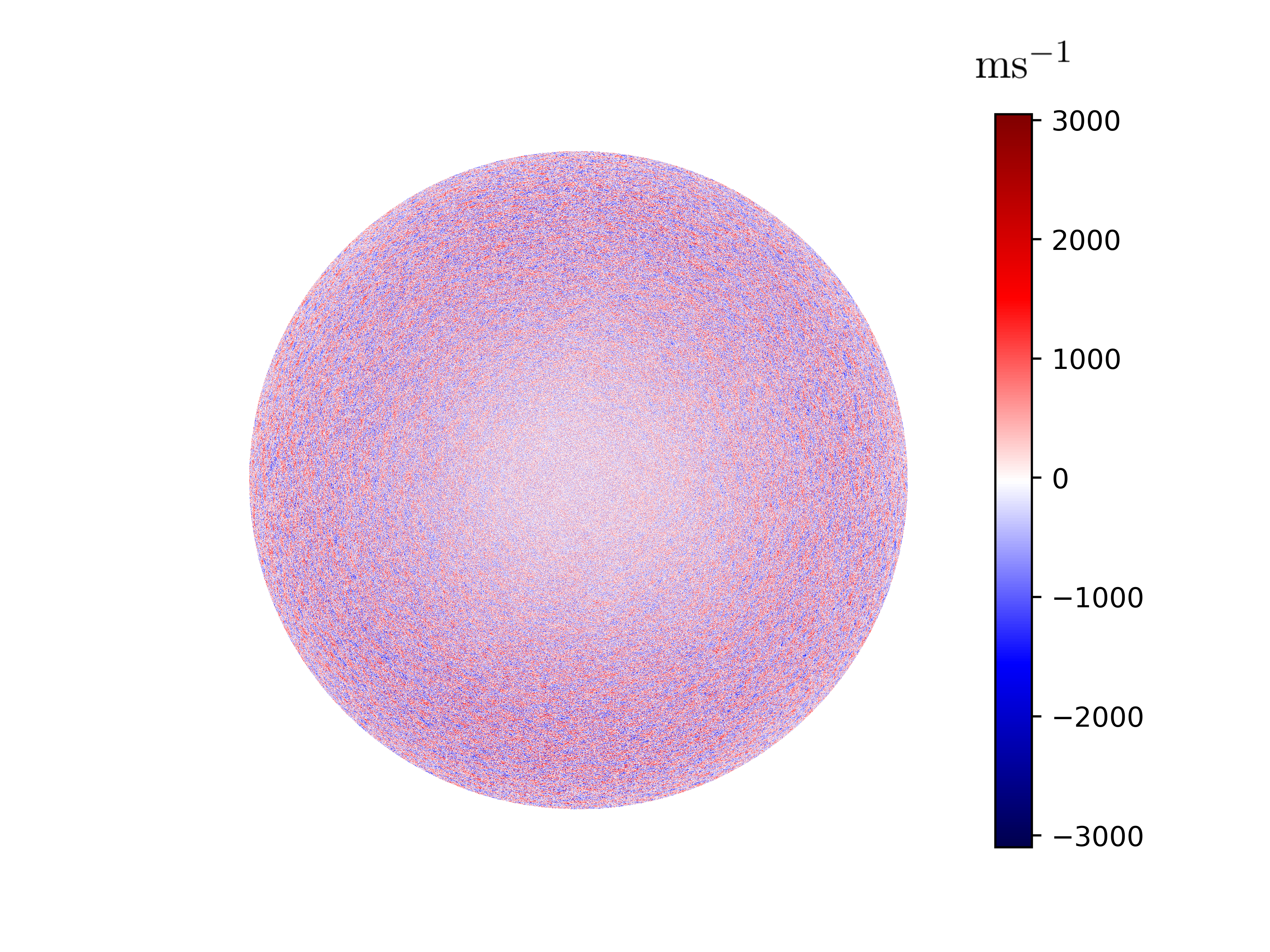}
\caption{Residual map after the removal of rotation, meridional circulation 
and limb shift. The disk center is lighter than the limb indicating very 
low radial velocity component when compared to horizontal velocity.}
\label{fig:residualMap}
\end{figure}

% imaging artefacts
The final effect that needs to be removed is the imaging artefact. This appears in
the form of fringes on the solar disk. We obtain the artefact by averaging over
100 images and subtracting the mean.
After removing all these effects, the residual map, 
shown in Figure~\ref{fig:residualMap}, is ready for analysis.

%\tb{Lindegren \& Dravins, 2003} -

\newpage
%------------------------begin sec:data-analysis--------------------------------
\section{Data Analysis}\label{sec:data-analysis}
\begin{figure}[!ht]
\centering
\includegraphics[width=0.3\textwidth]{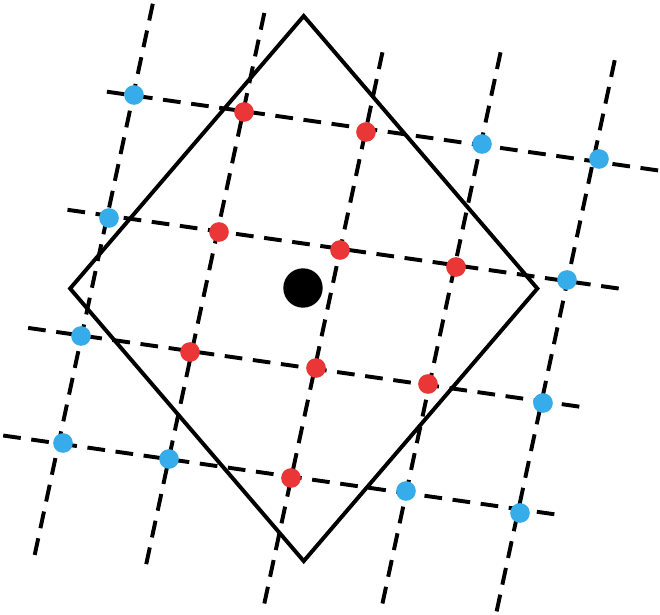}
\caption{The boundary of a HEALPix pixel is shown in solid line with its coordinates
indicated by black dots. The HMI grid points are shown using dashed lines. The grid points
which lie within the chosen HEALPix pixel are shown in red and the points which lie outside
the pixel are in blue.}
\label{fig:healpy-hmi}
\end{figure}

We perform a spectral analysis on the preprocessed data  using the Python 
package \texttt{healpy} \citep{healpy}. This is based on the HEALPix scheme. The HEALPix scheme
divides the surface of a sphere into $N_{\text{pix}}$ pixels of equal area. It also
enables an accurate representation of all signals with $\ell \le \ell_{\text{max}}$, where
$\ell_{\text{max}}$ depends on $N_{\text{pix}}$. Each pixel is denoted by its coordinates
at its centre. Hence, these pixels are different from grid points on which HMI Dopplergram
data are recorded. A schematic overlay of a HEALPix pixel and grid points from HMI 
is shown in Figure~\ref{fig:healpy-hmi}. Motivated by radial-velocity measurements of the
Sun, which are obtained using the disk-integrated velocity \citep{wright-kanodia-2020}, we
use an integrated velocity to determine the effective Doppler velocity of the pixel,
\begin{equation}
 v^{\text{hp}} A_{\text{pix}} = \int_{A_\text{pix}} dA \, \, v^{\text{HMI}}
 \approx \sum_i dA_i v_i^{\text{HMI}},
\end{equation}
where $v^{\text{hp}}$ is the velocity of the HEALPix pixel, 
$v^{\text{HMI}}$ is the observed velocity from HMI, 
$A_{\text{pix}}$ is the area of HEALPix pixel and 
the integration is performed over the region of a HEALPix pixel. 
Repeating this for each pixel, we obtain a HEALPix map 
corresponding to the HMI Dopplergram image. 
This map is used to obtain the observed spectral 
parameters $(u_{st}, v_{st}, w_{st})$.
The linear relation in Eqn.~(\ref{eqn:mixing-matrix-compact}) 
may be used to determine the ``true'' spectral parameters $\tilde{\mathcal{U}}$ by inverting mixing-matrix 
$\mathcal{M}$ in order to compute $\tilde{\mathcal{U}}$. 
For $\ell_{\text{max}} = 1500$, $\mathcal{M}$
has a size $3,377,250 \times 3,377,250$. 
This large matrix may be broken down into a number of 
smaller dimensional sub-matrices, making the inversions 
more tractable. 
The coordinate transformation is shown for 
$\mathcal{M}^{(uv)\ell m}_{st}$, but it holds for all the 
sub-matrices given in Table~(\ref{tab:mixlist}). 
The transformation involves moving the pole to disk-center, 
which enables integration over the entire domain of the
azimuthal coordinate $\phi$. 
Since all the VSH components are separable functions
in $\theta$ and $\phi$, with the dependence on $\phi$ 
appearing as a phase factor $\exp(im\phi)$,
where $m$ is the azimuthal quantum number, 
the sub-matrices reduce to a function of only the
angular degrees $s, \ell$.

\begin{align}
 \mathcal{M}^{(uv)st}_{\ell m} & = \int_0^{\pi} \sin\theta d\theta W(\theta) 
 \int_0^{2\pi} d\phi \, \left( \los\cdot \vec{\Psi}_{st}\right) 
 \left( \los \cdot \vec{Y}^*_{\ell m}\right) \nonumber \\
 & = \int_0^\pi \sin\theta \, d\theta \, W(\theta) f^s_\ell(\theta) \, 
 \int_0^{2\pi} d\phi \exp(i(m-t)\phi) \nonumber \\
 & = \delta_{m}^t \int_0^\pi \sin\theta \, d\theta \, W(\theta)\, f_\ell^s(\theta),
\end{align}
where $f_\ell^s(\theta)$ is the $\theta$-dependent part of 
\(\left( \los\cdot \vec{\Psi}_{\ell m} \right) 
  \left( \los \cdot \vec{Y}^*_{st}\right)\). 
It is seen that, in this coordinate system, 
there is no mixing between different azimuthal quantum numbers. Hence, we may separate the linear equation
given by Eqn.~(\ref{eqn:mixing-matrix-compact}) into independent equations for each $0 \le m \le \ell_{\text{max}}$.
Since the mixing atrices for each azimuthal order are independent of each other, their 
inverses are computed in a parallel fashion by using \texttt{GNU Parallel} by \cite{gnu-parallel}. These
matrices have very large null spaces; the mixing-matrices transform the true spectral parameters in to 
the observed spectral parameters and hence there is a loss of information due to observing only
the line-of-sight component of half the solar surface. 

The regularized inverse is computed 
by using unit regularization. 
\begin{equation}
 \tilde{\mathcal{U}} = \left( \mathcal{M}^T \mathcal{M} + \lambda I \right)^{-1} \mathcal{M}^T \mathcal{U},
\end{equation}
where $\lambda$ is the regularization parameter.

A similar analysis by \cite{rincon-2017} employed
CST to obtain the velocity spectrum. At the outset, we note 
that CST generates robust estimates
of horizontal velocities. However, since the velocity is 
measured by tracking granular structures, the spectrum 
is limited to $\ell < 850$, whereas the current work 
allows us to image up to $\ell = 1535$. The current work can be extended to
even higher $\ell$, provided sufficiently resolved
observations are available.

\section{Goodness of inversion}\label{sec:goodness-inversion}
The mixing matrix $\mathcal{M}$ transforms the complete velocity 
spectrum on the entire surface of the Sun to the line-of-sight spectrum
on less than half the solar surface. Hence, $\mathcal{M}$ has a 
large null-space; the observed spectrum is insensitive to 
(a) components of velocity perpendicular to the line of sight, and
(b) all components of velocity on the farside of the Sun. 
To understand how well the inversion is able to reproduce the 
true spectra, we construct a variety of qualitatively 
different synthetic spectra given below. 

\begin{figure}[!ht]
    \centering
    \includegraphics[width=\textwidth]{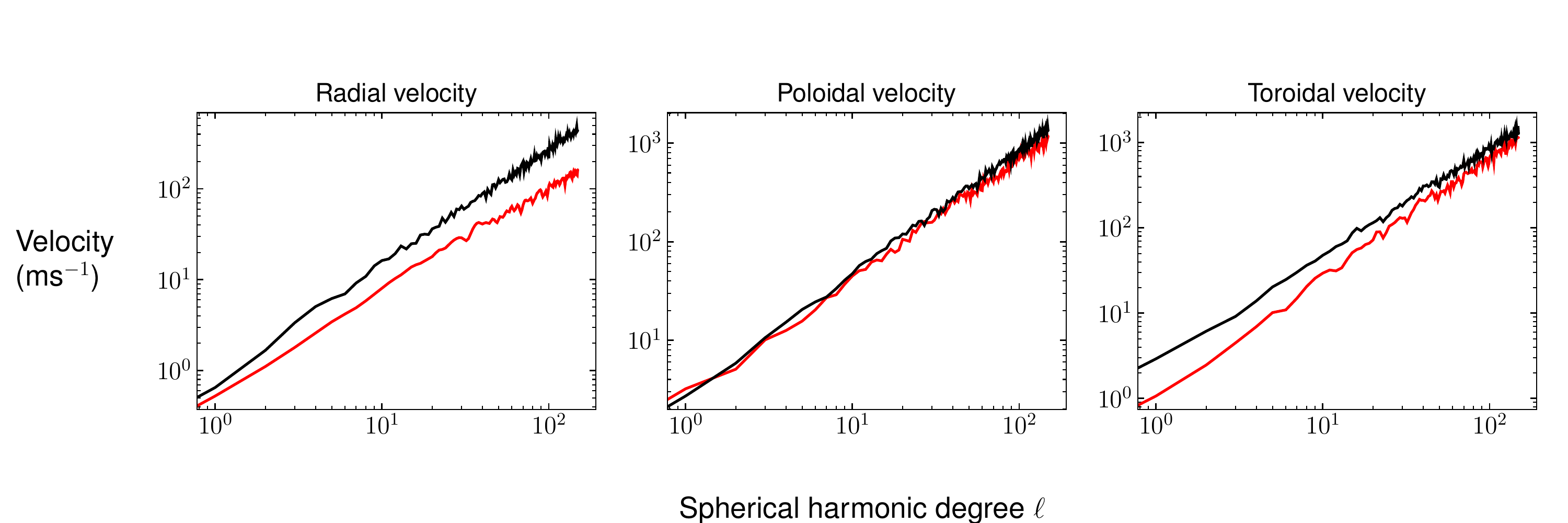}
    \caption{Comparison of actual (black) and inverted (red) 
    velocity profiles for sectorally dominated synthetic spectra. 
    The poloidal spectrum reproduction is significantly 
    superior to the radial and toroidal spectra.}
    \label{fig:synth-sectoral}
\end{figure}

For sake of illustration, only synthetic tests for sectoral 
harmonics are shown in Figure~\ref{fig:synth-sectoral}. 
It is seen that the inversion of the poloidal component is 
the most accurate. Inverted values of radial and toroidal 
components are seen to follow the same power law, 
but differ in magnitude when compared to the true spectrum. 
The inversion tests for all the other cases are presented in 
Appendix~\ref{appendix:synth-test}. 
Across all the synthetic profiles, the inversion of 
the poloidal component seems to be accurate and hence robust.
It is also observed that the inverted spectra of all 
flow components are systematically
underestimated in the synthetic tests.

\section{Results and Discussion}\label{sec:results-discussion}
The HEALPix method involves constructing and heirarchically subdividing
the entire spherical surface. The number of subdivisions is denoted 
by $n_d$ and the corresponding number of HEALPix pixels is given by
$12\times 4^{n_d}$, i.e., $n_d = 0$ corresponds to the largest pixels 
where the sphere is equally divided into 12 pixels. All subsequent 
refinement in pixel size involves dividing cells into 
4 sub-divided cells and hence the number of pixels $N_\text{pix} = \{12, 48, 192 ...\}$.
Mapping the observed HMI Dopplergram to HEALPix involves
populating the pixels with observed data. 
Using a very high resolution would result in a lot of ``holes'' 
in the remapped data. This happens when the pixel size using HEALPix
is smaller than the existing grid. A very low resolution 
would result in loss of information at smaller spatial scales and hence
not preferred. For the 16 megapixel data from HMI, it was found 
that the optimal $n_d = 9$ ($N_\text{side} = 512$ in HEALPix terminology).
For this resolution, HEALPix may accurately represent band-limited 
signals up to spherical harmonic degree $\ell = 1535$. HEALPix also enables faster computation of spectrum as the 
computation of spherical harmonic transform scales as 
$\mathcal{O}(N_\text{pix}^{1/2})$,
thus making it attractive for processing of high-resolution data.

\begin{figure}[!ht]
 \centering
 \includegraphics[width=\textwidth]{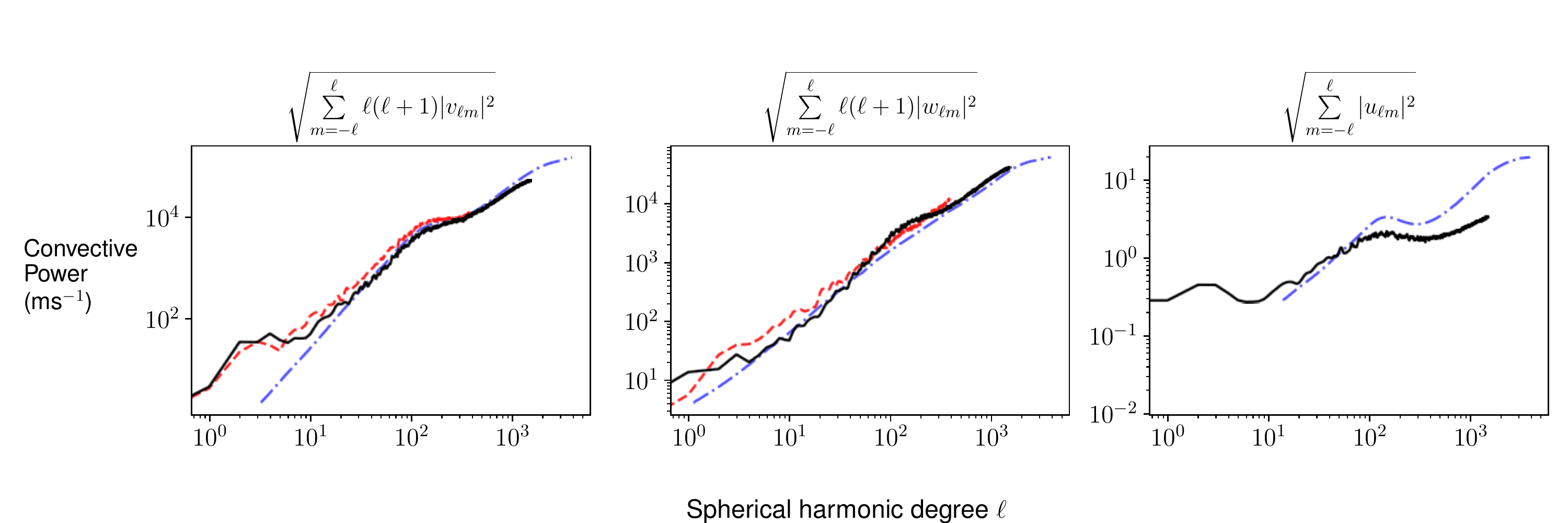}
 \caption{Comparison of convective power spectrum from 3 different methods. 
 The present work is indicated by solid black line. 
 The spectrum obtain from LCT is shown as a dashed red line. 
 The spectrum from H15 is shown in dash-dotted blue line. The present work agrees with inferences from LCT better than H15 for poloidal flow. LCT measures only horizontal velocity
 and hence the radial flow comparison is with H15 only.}
 \label{fig:compare3}
\end{figure}

The mixing matrix is constructed for all $ m \le 1535$ and 
inversion is performed to obtain the full spectrum.
We compare the results of inversion from forward modelling 
of H15 and spectrum obtained from LCT in Figure~\ref{fig:compare3}. Supergranules have finite depth extents and are sensitive to flows at a range of sub-surface layers. The estimates of flow velocities that LCT (which uses supergranular motions) provides are thus likely to be different from those derived using Dopplergrams because the depth averaging in these two measurements is different. We would thus expect qualitative and approximate quantitative agreement between the two techniques.

While H15 estimates spectra for $\ell \le 4096$, the current work 
is limited to $\ell \le 1535$ and LCT is limited to $\ell \le 383$.
The poloidal flow shows a clear peak at $\ell \approx 120$ which 
is the well known peak of supergranulation. 
This peak is seen in the spectrum obtained from LCT as well as H15.
Note that, although LCT and H15 are in good agreement over the range 
$60 \lessapprox \ell \lessapprox 120$, they qualitatively differ for 
$\ell \lessapprox 60$. 
On the other hand, inferences from the current work 
are in good qualitative agreement with LCT, 
but the magnitudes are different. 

As evident from the synthetic tests, the magnitude is an 
underestimate of the true convective amplitudes. 
Since supergranulation is dominantly poloidal, the peak near 
$\ell=120$ should appear only in the radial and poloidal components. 
However, it is seen that the supergranulation 
peak appears in the toroidal flow inversion as well. 
This is due to mode mixing and the inability of the 
inversion to completely separate out the 
poloidal from the toroidal component. 
The ability of the inversion to distinguish between the 
magnitudes of radial and poloidal flows hints at 
limited contamination of radial flows by poloidal flows.

\begin{figure}[!ht]
 \centering
 \includegraphics[width=0.6\textwidth]{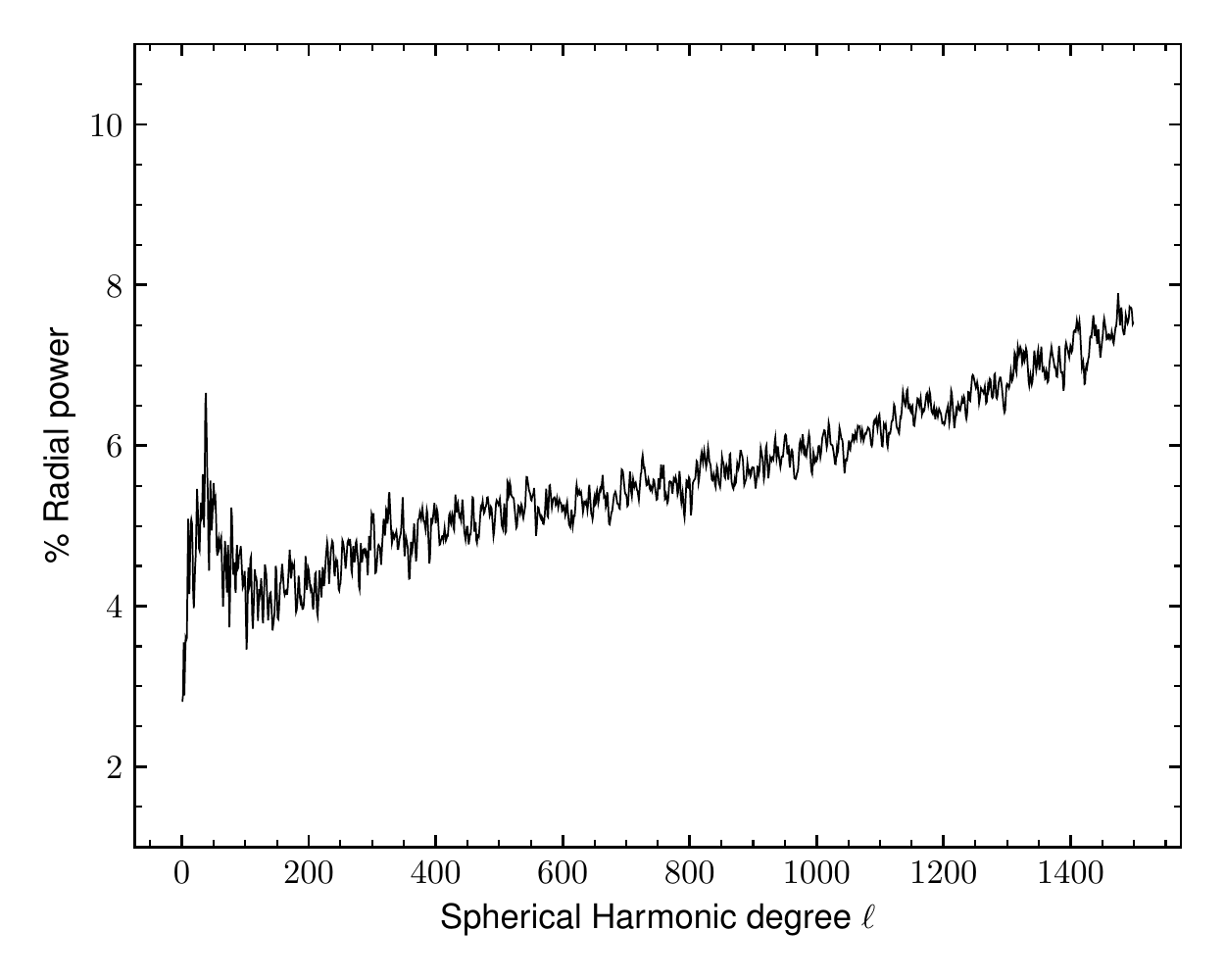}
 \caption{Fraction of total power contained in radial component as a function of 
 spherical harmonic degree. The fraction increases with wavenumber.
 A supergranular peak at around $s=120$ is also seen here, hinting at its convective origin.
 Although the peak could be due to the contamination from the 
 poloidal power, synthetic tests show that contamination is minimal
 (Appendix~\ref{sec:contamination}). The radial
 flow magnitude agrees more with the measurement of vertical RMS velocity
 \cite{duvall-birch-2010} than other measurements.}
 \label{fig:rad_power}
\end{figure}

Radial power increases with angular degree, 
with spatial scales near granulation having the highest power, since granulation comprises strong 
upflows and downflows.
The magnitude of the radial component is also underestimated
when compared with H15, although the fractional radial 
convective power is in good agreement with previous 
studies \citep{hathaway2002,hathaway2015}.  As shown in 
Figure~\ref{fig:rad_power}, at lower $\ell$, the radial power 
only corresponds to $5\%$ of the total power and 
at $\ell \approx 1500$, it contributes almost
$8\%$. The peak radial velocity  at supergranular scales
is found to be 2.2 ms${}^{-1}$,
which is an underestimate when compared to \cite{duvall-birch-2010} 
(4 ms${}^{-1}$), \cite{hathaway2002} (13 ms${}^{-1}$) and measurement of 
20 ms${}^{-1}$ by \cite{rincon-2017}. It is also within the upper limit of  
10 ms${}^{-1}$ set by \cite{giovanelli1980}.

To study the temporal characterisitcs of the inverted
convective flow spectra, we process 1 year's worth of data
at a rate of 1 image per day, with a maximum (Nyquist) frequency of 
$5.6\, \mu$ Hz and a frequency bin size of $31.7$ nHz.
We compute the convective power in all the convective flow 
components, as a function of spherical harmonic degree $\ell$ and 
temporal frequency $\nu$. However, the contour plot of $\ell-\nu$ is
not informative, i.e., we do not observe clear 
characteristic frequencies in radial, poloidal or 
toroidal flow components.
%It is 
%seen in Figs.~(\ref{fig:v-freqpow}, \ref{fig:w-freqpow}) that %neither
%toroidal nor poloidal components of flow vary with 
%characteristic frequencies, indicated by the flat profiles as 
%a function of frequency. 
%However, processing of LCT reveals a peak in frequency near 
%$1$ $\mu$Hz range for poloidal flows for very large length
%scales to the supergranular length scale. 
Other studies of convection have observed supergranular waves which are 
found to have oscillation periods of $\sim 2$ $\mu$Hz 
\citep{gizon-2003, langfellner2018}.
Additionally \cite{hanasoge2020} estimated that the  toroidal flow power 
increased significantly with temporal frequency.

%however, the
%results in Figure~\ref{fig:w-freqpow} show that this power is %nearly constant
%with frequency. The toroidal flow does not show temporal
%variation in either LCT or the present work, except at very
%low wavenumbers. 
%A more detailed comparison of LCT and 
%inversion is presented in Figs.~(\ref{fig:figv-sigma},
%\ref{fig:figw-sigma}) of Appendix~\ref{sec:comparison-lct}.

%, as shown in
%Figure~\ref{fig:vsigma-lct-compare}

\begin{figure}[!ht]
 \centering
 \includegraphics[width=\textwidth]{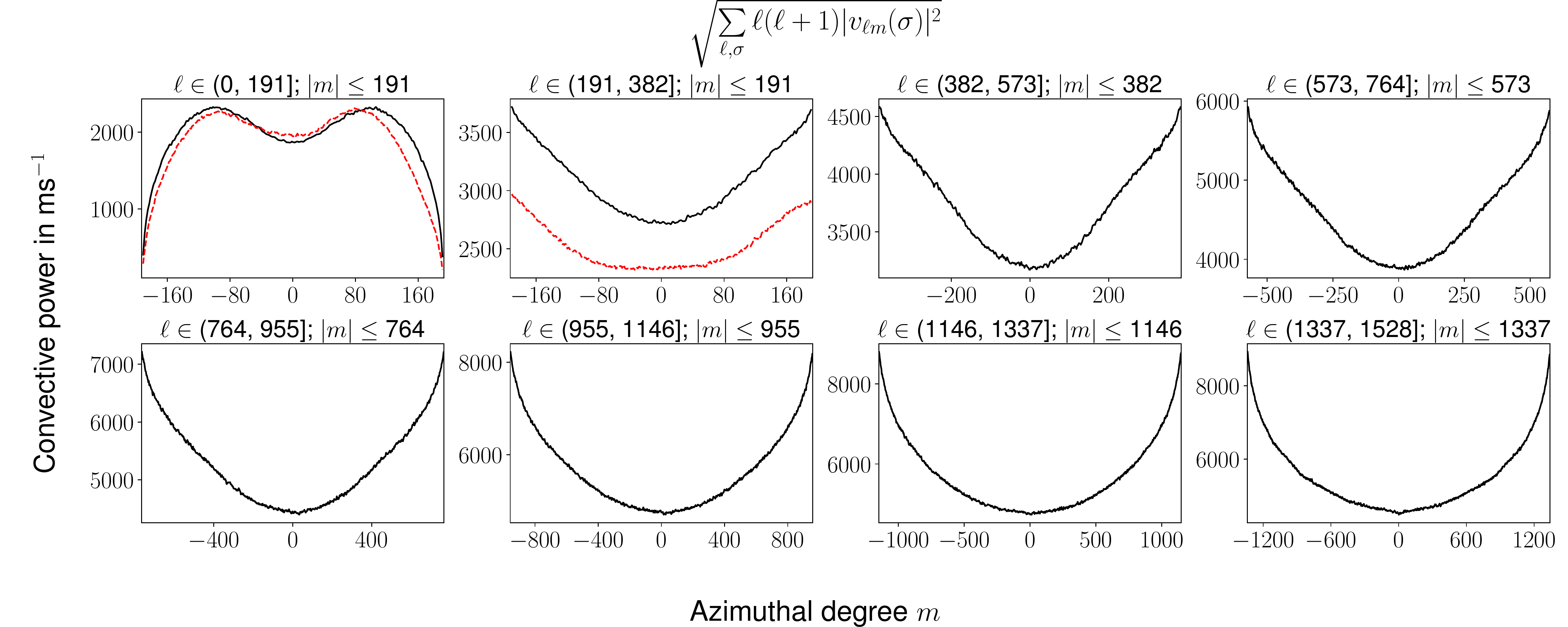}
 \caption{Distribution of poloidal flow power as a function of azimuthal order:
 LCT data is in red and the present work is shown in black.
 The power falling at the largest $m$ are due to fewer 
 modes being summed.}
 \label{fig:v-allpow}
\end{figure}

\begin{figure}[!ht]
 \centering
 \includegraphics[width=\textwidth]{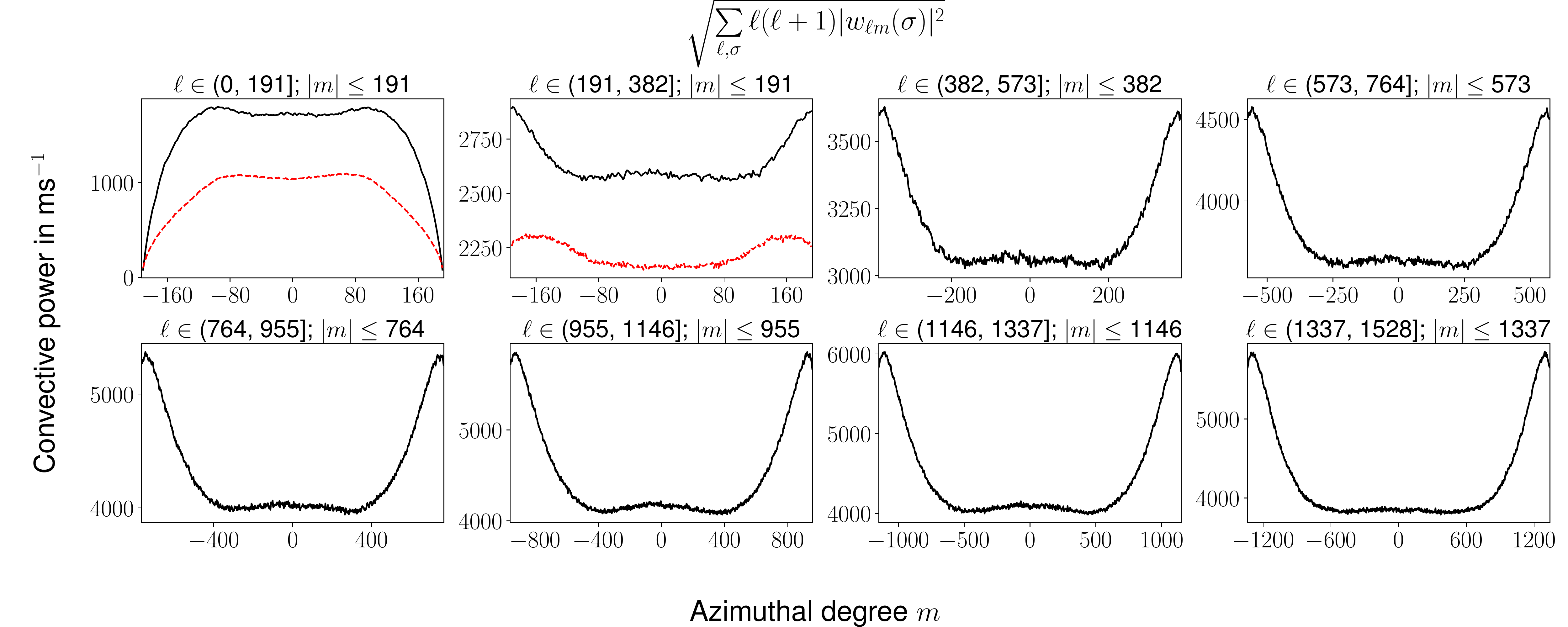}
 \caption{Distribution of toroidal flow power as a function of azimuthal order. 
 LCT data is shown in red and the present work is shown in black.
 Power declining at the highest $m$ is attributed to fewer 
 modes being summed. LCT and inversion curves are different in both 
 magnitude and trend.}
 \label{fig:w-allpow}
\end{figure}

\begin{figure}[!ht]
 \centering
 \includegraphics[width=\textwidth]{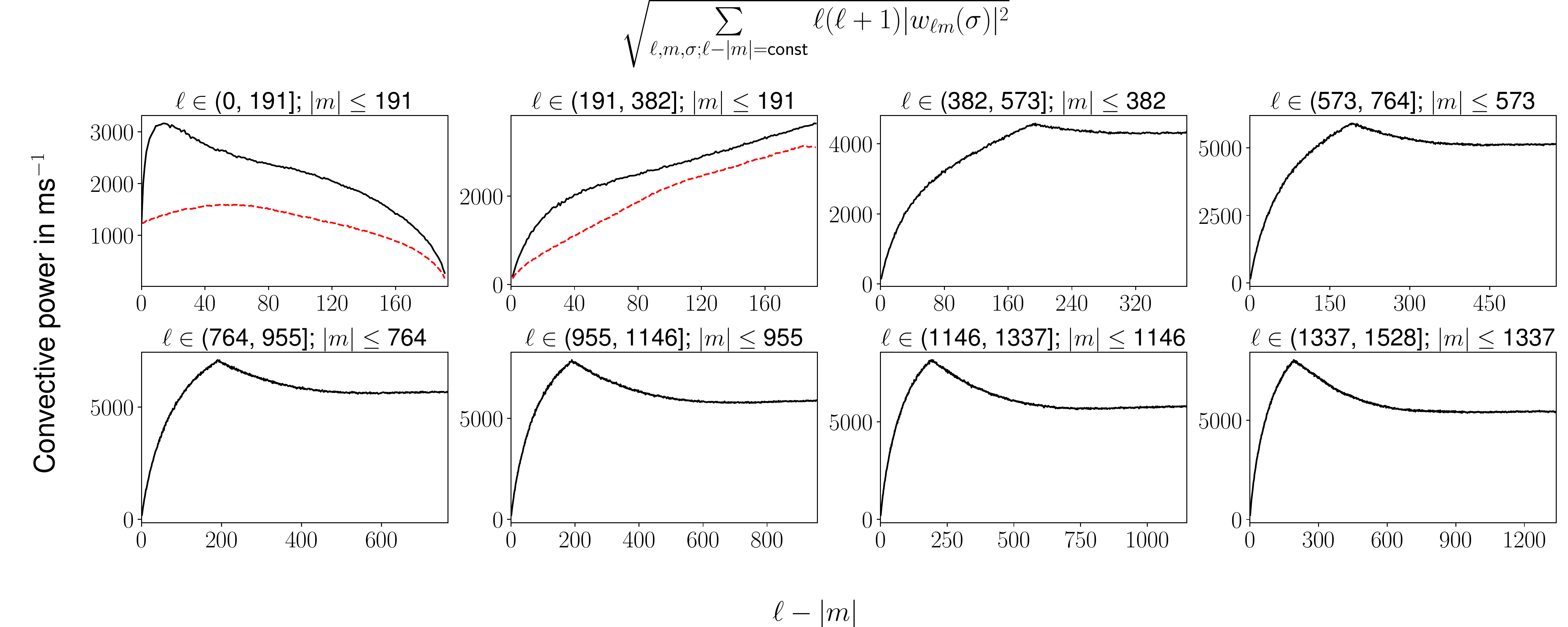}
 \caption{Distribution of toroidal flow power as inferred 
from LCT (red) and the current work (black). The profiles from inversion
peak at low $\ell-|m|$ whereas LCT data attain their maximum at 
intermediate $\ell-|m|$.}
\label{fig:compare-figw-s-|t|}
\end{figure}

Inversions for the full spectrum enable us to characterise the 
shapes of convection at different length scales. Sectoral
harmonics show power concentrated at low latitudes. Tesseral harmonics have 
power distributed across the entire surface and zonal harmonics
have axisymmetric structure. This classification can
be quantified using $\ell-|m|$, which quantifies the number of
nodes in latitude. 
We plot the distribution of power as a function of $\ell - |m|$ 
for poloidal flows in Figure~\ref{fig:v-allpow} and for 
toroidal flows in Figure~\ref{fig:w-allpow}. 
Irrespective of the range of $\ell$, we observe that the 
poloidal convective power is maximum for low $\ell - |m|$ 
i.e. sectoral harmonics. This is also seen to be in qualitative
agreement with LCT data, as seen in Figure~\ref{fig:compare-figw-s-|t|},
except at very low $\ell-|m|$. A more detailed comparison is 
presented in Figs.~(\ref{fig:figv-s-|t|}, \ref{fig:figw-s-|t|})
of Appendix~\ref{sec:comparison-lct}.
%Thermal flux is carried by poloidal flow, 
%whereas toroidal flow, which by construction is divergence free, cannot directly transport heat. 

\begin{figure}[!ht]
\centering
\includegraphics[width=\textwidth]{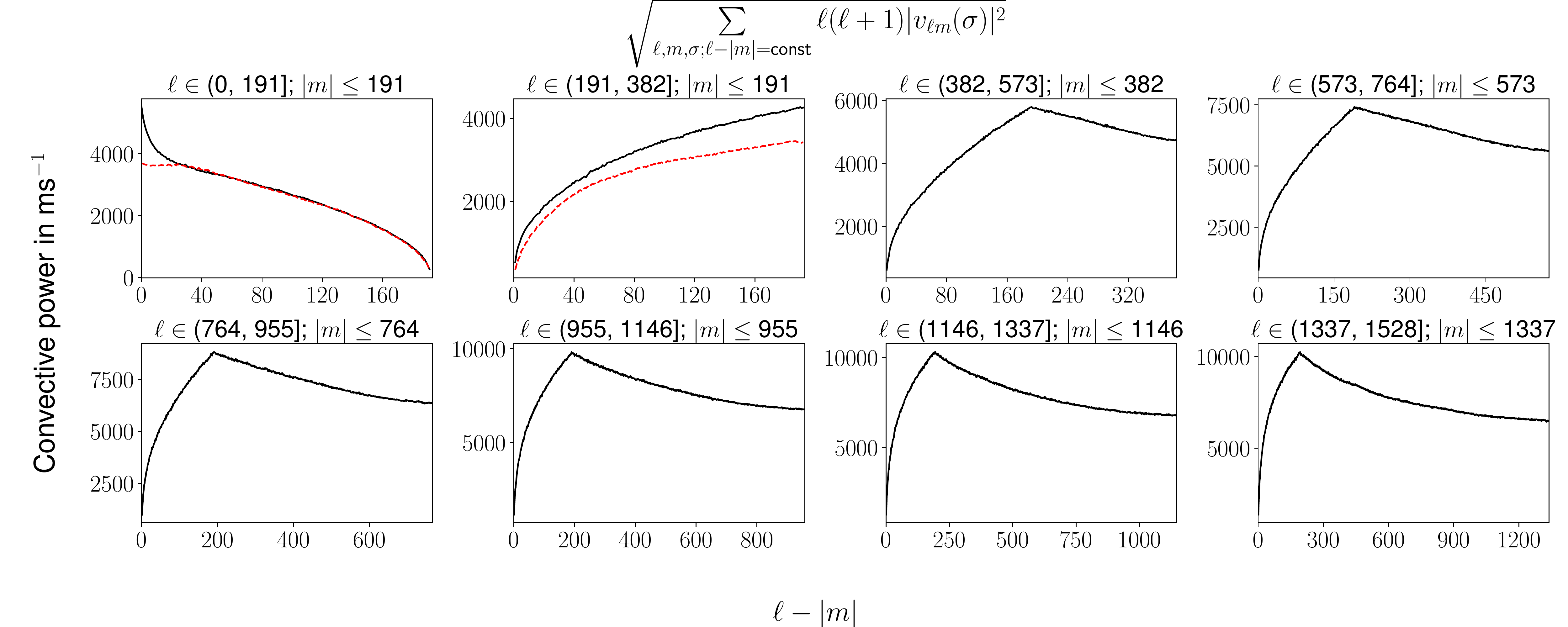}
\caption{Distribution of toroidal flow power as inferred 
from LCT (red) and the current work (black).  A strong peak is observed 
for inversions at very low $\ell-|m|$ which isn't seen in LCT data.}
\label{fig:compare-figv-s-|t|}
\end{figure}

{
Solar differential rotation is known to have conical 
(as opposed to cylindrical) isorotation contours inside the 
convection zone \citep{gilman-howe-2003}.
%{\bf REFERENCE}. 
Theoretical modelling \citep{balbus2009} suggests that 
a latitudinal temperature gradient is necessary both in the convection zone
and at the tachocline to sustain the observed rotational shear.
Non-cylindrical differential
rotation is driven by a combination of Reynolds stresses and a latitudinal temperature gradient 
\citep{miesch2005lsrp}.
Measurements of surface temperatures
have shown latitudinal variations of a few K 
\citep{kuhn1998, rast2008} with a  
minimum at mid-latitudes.  A temperature minimum at mid-latitudes
also hints at preferred latitudes for convection and
could potentially be the reason for the peak at finite $\ell - |m|$ seen in the poloidal
flows shown in Figure~\ref{fig:compare-figv-s-|t|}.}

%\begin{figure}[!ht]
% \centering
% \includegraphics[width=\textwidth]{compare_figv_sigma.pdf}
% \caption{Poloidal flow power as function of temporal frequency. 
% Power obtained from LCT (red) is limited to $\ell = 383$, but is 
% truncated at $\ell=374$ for ease of plotting. Power computed in %the
% present work (black) shows very weak dependence on temporal %frequency; the variation
% is less than 15\% at the lowest frequencies and roughly constant %thereafter.}
% \label{fig:v-freqpow}
%\end{figure}

%\begin{figure}[!ht]
% \centering
 %\includegraphics[width=\textwidth]{w_freqpow.pdf}
% \includegraphics[width=\textwidth]{compare_figw_sigma.pdf}
%  \caption{Toroidal flow power as function of temporal frequency. 
% Power obtained from LCT (red) is limited to $\ell = 383$, but is 
% truncated at $\ell=374$ for ease of plotting. Power computed in %the
% present work (black) shows very weak dependence on temporal %frequency; the variation
% is less than 15\% at the lowest frequencies and roughly constant %thereafter.}
% \label{fig:w-freqpow}
%\end{figure}

Seismic analyses by \cite{hanasoge2020} indicate that toroidal power peaks 
at the equator and is very small beyond $\ell-|m| > 10$, for $\ell < 50$. 
Our inversions here for toroidal flow power attain a maximum around 
$\ell - |m| \sim 20$ (Figure~\ref{fig:w-allpow}) whereas LCT data 
shows a dominant peak at $\ell - |m| \sim 100$. 
However, given the poor reproduction of 
toroidal spectrum in the synthetic tests, the results of 
inverted toroidal flows are not reliable. 
%Improvements to inferring 
%the true spectrum could be sought by performing constrained %inversions. 
%The toroidal flow being divergence-free, 
%a constrained inversion would look for solutions which %minimizes the 
%mixing of toroidal and poloidal spectra. {\bf YOURE ALREADY CONSTRAINING IT}

Understanding the origin of supergranulation has been
a long-standing challenge \citep{rieutord2010}. 
`Realistic' numerical simulations (which account for 
ionization and radiative transfer) have been successful at reproducing
granular scales \citep{stein-nordlund-1998}. 
While \cite{rast2003} attributed supergranulation to be 
emergent from advective interaction of granular plumes,
attempts at simulations at these scales have been made through
two different approaches 
(a) global spherical simulations where supergranulation 
corresonds to the smallest scale \citep[e.g][]{derosa-gilman-toomre-02}
and (b) local cartesian simulations where supergranulation is the
largest scale \citep[e.g.][]{cattaneo2001}. 
\cite{schrijver1997} suggested that granular and supergranular
have very similar features (after accounting for length-scale differences);
the real utility in inversion-based observational
studies such as the present is that it provides full spectral information
of flows at these scales, useful for potentially detecting trends and benchmarking numerical simulations and seismic analyses.

Experimental studies of forced rotating turbulence have 
shown the presence of a double turbulent cascade -- 
a direct cascade at smaller scales and an inverse cascade 
at larger scales \citep{campagne-2014}. This is similar to 
observed convective power around the the supergranular 
length scales. While \citet{rincon-2017} observe an 
$\ell^2$ scaling for convective
poloidal power for $\ell < 120$, we find that the scaling to be
closer to $\ell^4$, which is observed in homogeneous turbulence
\cite{batchelor-proudman-1956}; solar convection is known to be 
neither homogeneous nor isotropic \citep[owing to 
unstable stratification;][]{rincon-2007}. However,
the large null-space of the mixing-matrix prevents us from quantifying
the uncertainties in the inferred spectra and hence we are cautious about drawing detailed conclusions from these measurements.
The ratio of toroidal and poloidal flow power
can be used to identify the regime of turbulence that the Sun
operates in \citep[][experiments on Rayleigh-Bernard 
convection]{horn-shishkina-2015}, and the present work
enables the characterization of the horizontal
flows at very high $\ell$. 
The convective flow
power is found to flatten out at higher angular degrees 
near the granular length scales, and 
the present work would also be useful for the study
of convection at sub-granular length scales, when
higher resolution observations become available.\\

Some of the results in this paper have been derived using the healpy and HEALPix packages. SGK thanks Pranav Sampathkumar (Karlsruhe Institute of Technology) for help with HEALPix.
LCT data was generated by Bjoern Loeptien (Max Planck Institute for Solar System Research) and provided by 
the authors of \cite{hanson2020}. SMH thanks R. Bogart (Stanford University) and D. Hathaway (NASA Ames) for useful conversation.
{The authors thank Aimee Norton (Stanford Univeristy)
for help with HMI data.}  The authors also thank 
the anonymous referee for valuable suggestions that 
helped improve the text in this  manuscript.

\clearpage
\appendix
%\counterwithin{figure}{section}
\section{Vector Spherical Harmonics} \label{sec:vsh}
The vector spherical harmonic components are given by
\begin{align}
 \vec{Y}_{lm}(\theta, \phi) & \equiv \hat{\vecr} Y_{lm}(\theta, \phi) \\
 \vec{\Psi}_{lm}(\theta, \phi) & \equiv \vec{\nabla}_h Y_{lm}(\theta, \phi) \\
 \vec{\Phi}_{lm}(\theta, \phi) & \equiv \hat{\vecr} \times
                                \vec{\nabla}_h Y_{lm}(\theta, \phi),
\end{align}
where $Y_{lm}(\theta, \phi)$ are spherical harmonics and $\hat{\vecr}$ is the
radial unit vector and $\vec{\nabla}_h$ is the horizontal gradient operator
given by
\begin{equation}
 \vec{\nabla}_h = \hat{\theta}\frac{\partial}{\partial \theta}
                + \hat{\phi}\frac{1}{\sin\theta}\frac{\partial }
                {\partial \phi}.
\end{equation}
The vector spherical harmonics are orthogonal. The orthonormality
may be expressed in compact form if we define \(\Lambda^0_{lm}
\equiv \vec{Y}_{lm}\), \(\Lambda^1_{lm} = \vec{\Psi}_{lm}\) and
\(\Lambda^2_{lm} = \vec{\Phi}_{lm}\).

\begin{align}
  \int d\Omega \vec{\Lambda^i}_{lm} \cdot \vec{\Lambda^j}^*_{l'm'}
  = N_{il} \delta^{ij} \delta_{ll'}\delta_{mm'},
\end{align}
where \(N_{il}\) is a normalization constant,
\(N_{0l} = 1, N_{1l} = N_{2l} = l(l+1)\) and
\(d\Omega = sin\theta d\theta d\phi\) is the surface element,
integration being performed over the entire surface of the Sun.

\section{Shifted Legendre Polynomials} \label{sec:legpoly}
The shifted Legendre polynomials are defined over 
the interval \([0, 1]\). In the current computation, we use shifted 
Legendre polynomials up to degree 5. They are defined as follow, 
\begin{align}\label{eqn:legpoly}
    P_0(x) & = 1 \nonumber\\
    P_1(x) & = 2x - 1\nonumber\\
    P_2(x) & = 6x^2 - 6x + 1\\
    P_3(x) & = 20x^3 -30x^2 + 12x -1\nonumber\\
    P_4(x) & = 70x^4 - 140x^3 + 90x^2 - 20x + 1.\nonumber
\end{align}
These polynomials are orthogonal and the orthogonality condition is 
given by
\begin{equation}
    \label{eqn:legpoly-ortho}
    \int_0^1 P_m(x) P_n(x) = \frac{1}{2n+1}\delta_{mn}.
\end{equation}

\section{Inversion -- Synthetic Test}
\label{appendix:synth-test}
\begin{figure}[!ht]
    \centering
    \includegraphics[width=\textwidth]{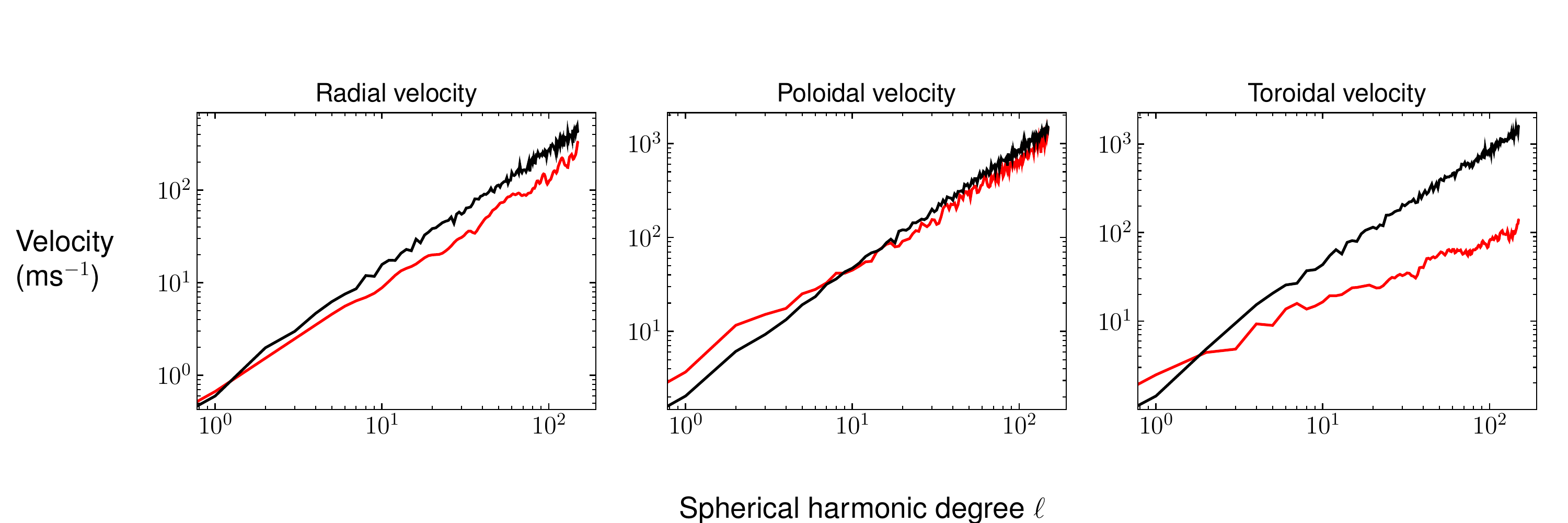}
    \caption{Comparison of actual (black) and inverted (red) 
    velocity profiles for a spectrum with zonal modes. The radial flows show the power law of the synthetic spectrum, but underestimates the
    magnitude. Poloidal inversion reasonably reproduces the synthetic
    spectrum for higher $\ell$. Toroidal inversion is erroneous.}
    \label{fig:synth-zonal}
\end{figure}

\begin{figure}[!ht]
    \centering
    \includegraphics[width=\textwidth]{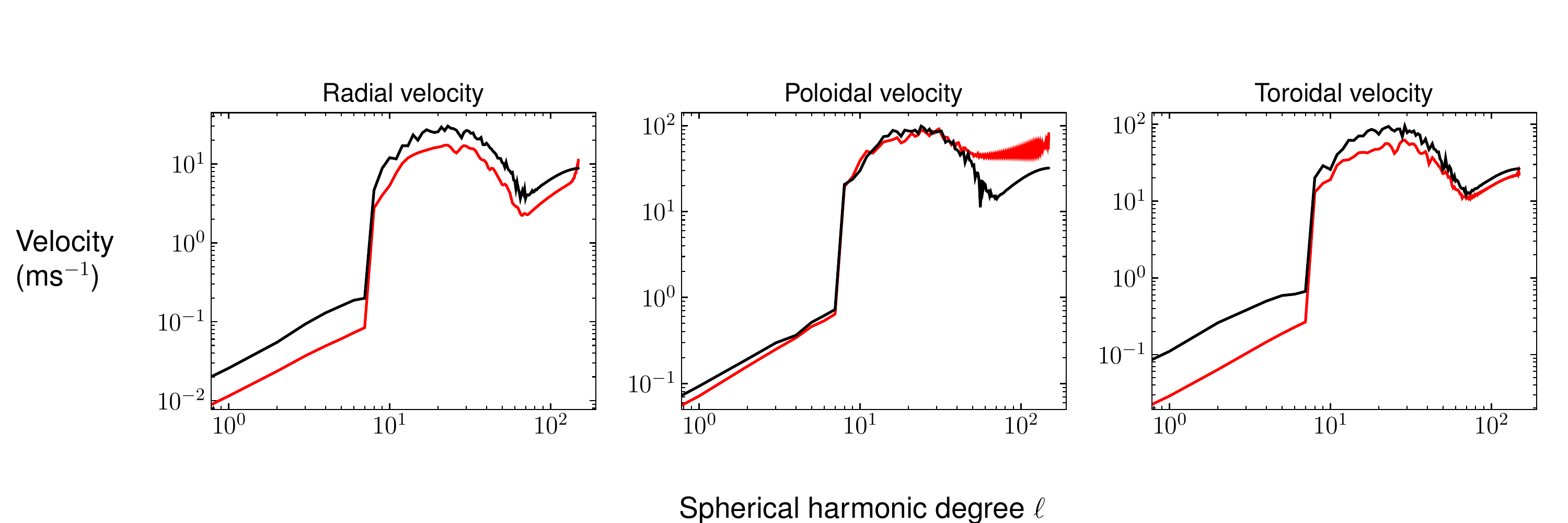}
    \caption{Comparison of actual (black) and inverted (red) 
    velocity profiles for a spectrum with tesseral modes. 
    Radial flows are underestimates of the input synthetic spectrum, whereas
    poloidal flows are reproduced except at high 
    $\ell$. Toroidal flows are underestimated.}
    \label{fig:synth-tesseral}
\end{figure}

\begin{figure}[!ht]
    \centering
    \includegraphics[width=\textwidth]{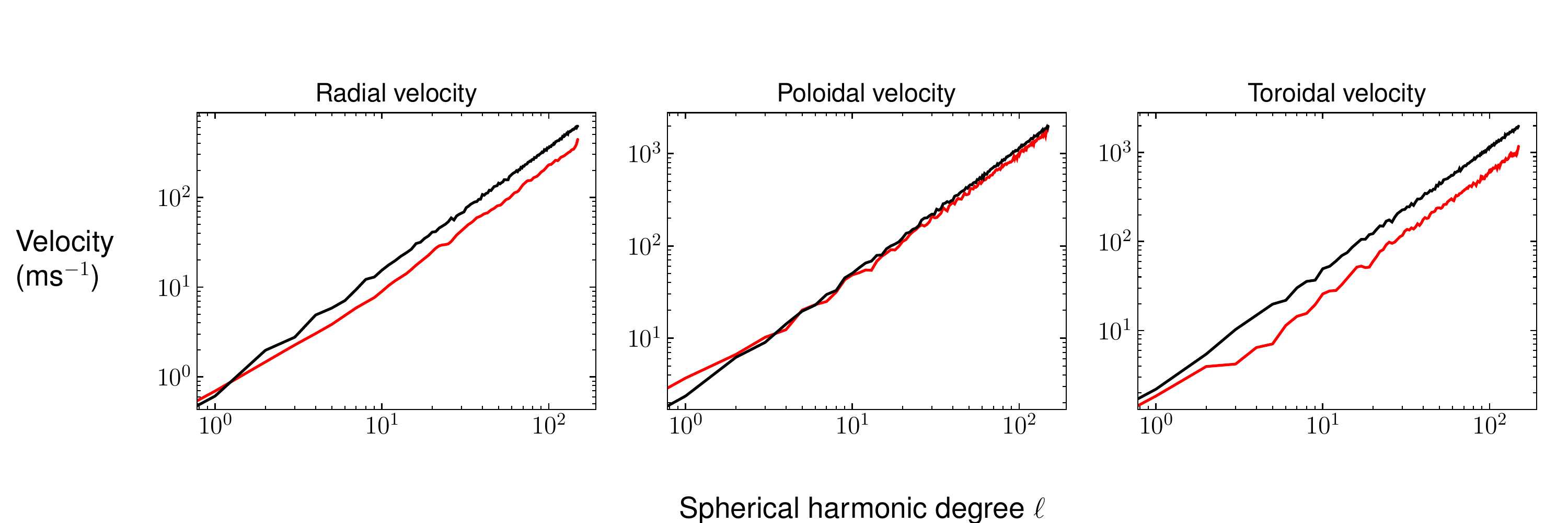}
    \caption{Comparison of actual (black) and inverted (red) 
    velocity profiles for a spectrum with all modes drawn from a 
    uniform random number. Poloidal flows are estimated correctly, 
    whereas radial and toroidal flows are underestimated.}
    \label{fig:synth-random}
\end{figure}

\begin{figure}[!ht]
    \centering
    \includegraphics[width=\textwidth]{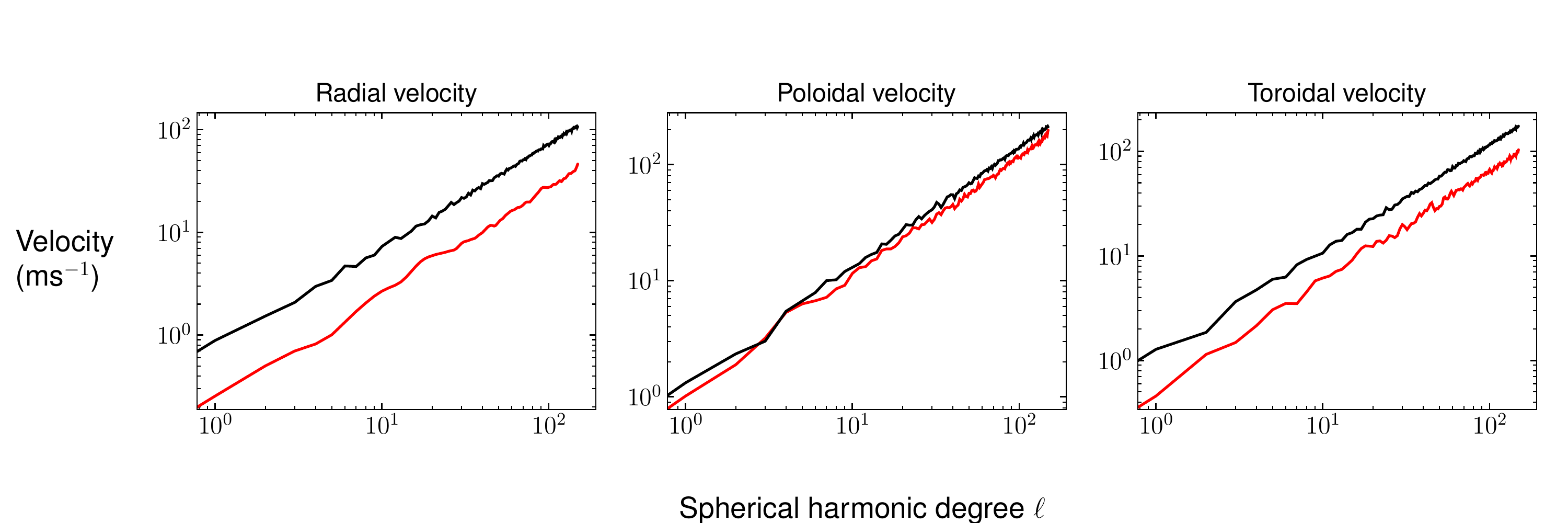}
    \caption{Comparison of actual (black) and inverted (red) 
    velocity profiles for spectrum with solar-like modes. Poloidal
    flows reproduction are the most accurate among the three.
    Radial flows are drastically underestimated, and toroidal flows
    are underestimated as well.}
    \label{fig:synth-solarlike}
\end{figure}

The mixing matrix \(\mathcal{M}\) needs to be inverted in order to obtain
the velocity components on the Sun \(U\) from the observed line-of-sight
velocity components \(\tilde{U}\) as \(U = \mathcal{M}^{-1} \tilde{U}\). 
The mixing matrix \(\mathcal{M}\) has a large null space and 
hence it becomes necessary to perform synthetic tests to 
establish how well we are able to reproduce 
different types of spectra.
 \begin{itemize}
 \item Zonal spectra (Figure~\ref{fig:synth-zonal})-- the synthetic spectra are non-zero only for $|t| \le 10$, where $t$ is the 
 azimuthal quantum number.
 \item Sectoral spectra (Figure~\ref{fig:synth-sectoral})-- the synthetic spectrum is non-zero only for 
 $s-|t| < 10$, where $s$ is the angular degree. 
 \item Tesseral spectra (Figure~\ref{fig:synth-tesseral})-- the synthetic spectrum is non-zero only for 
 $|t| \ge 10$ and $s-|t| \ge 10$.
 \item Random spectra (Figure~\ref{fig:synth-random})-- All spectral components are assigned a random number picked from a uniform
 distribution.
 \item Solar-like spectra (Figure~\ref{fig:synth-solarlike})-- Spectral components where radial components are small compared to horizontal component.
\end{itemize}
The goodness of inversion is summarized in Table~\ref{tab:inversion-summary}.
In all the cases, radial and toroidal flows are underestimated. The poloidal
flow inversion is the most accurate among the three. Zonal flow inversions
are the worst among all the cases, as the inversions capture neither
the magnitude of flow nor the power law as a function of wavenumber.

\begin{table}[!ht]
\centering
\begin{tabular}{|c|c|c|c|}
\hline
Spectrum type & Radial        & Poloidal & Toroidal \\
\hline
Zonal         & Underestimate & Good & Bad \\
Sectoral      & Underestimate & Good & Underestimate \\ 
Tesseral      & Underestimate & Good & Underestimate \\
Random        & Underestimate & Good & Underestimate \\
Solar-like    & Underestimate & Good & Underestimate \\
Sparse        & Underestimate & Good & Underestimate \\
\hline
\end{tabular}
\label{tab:inversion-summary}
\caption{Summary of goodness of spectral reproduction. 
The poloidal inversion works reasonably well for all
kinds of flows.}
\end{table}

\begin{figure}[!ht]
    \centering
    \includegraphics[width=\textwidth]{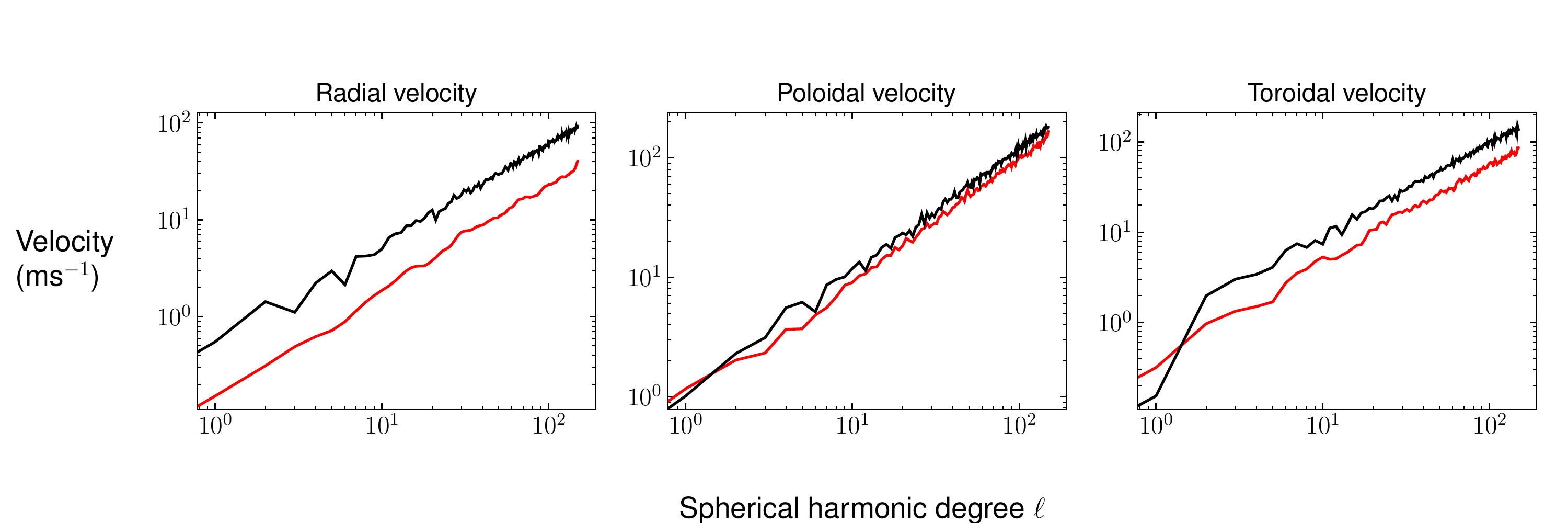}
    \caption{Comparison of actual (black) and inverted (red) 
    velocity profiles for spectrum with sparsely non-zero modes.
    Poloidal flow inversions work reasonably well. Radial and 
    toroidal flows are underestimated.}
    \label{fig:synth-sparse}
\end{figure}

\section{Comparison with LCT}
\label{sec:comparison-lct}
\begin{figure}[!ht]
 \centering
 \includegraphics[width=\textwidth]{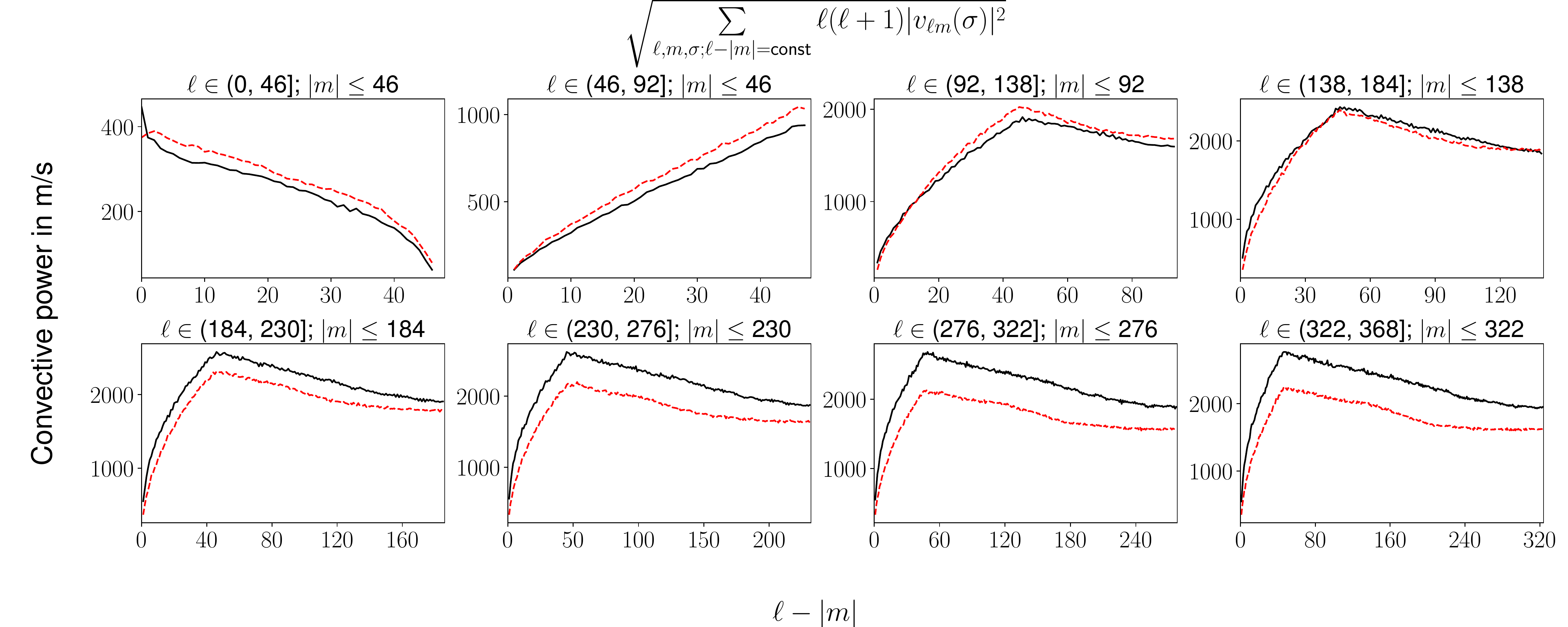}
 \caption{Distribution of poloidal flow power
 as a function of $\ell - |m|$, inferred 
 from LCT (dashed, red) and the current work (solid, black). LCT 
 indicates a preference for sectoral modes, but the power 
 falls off almost linearly as a function of $\ell-|m|$. Inversions
 show power concentrated at very low $\ell-|m|$, i.e., the strongest
 convection at low latitudes.}
\label{fig:figv-s-|t|}
\end{figure}

\begin{figure}[!ht]
 \centering
 \includegraphics[width=\textwidth]{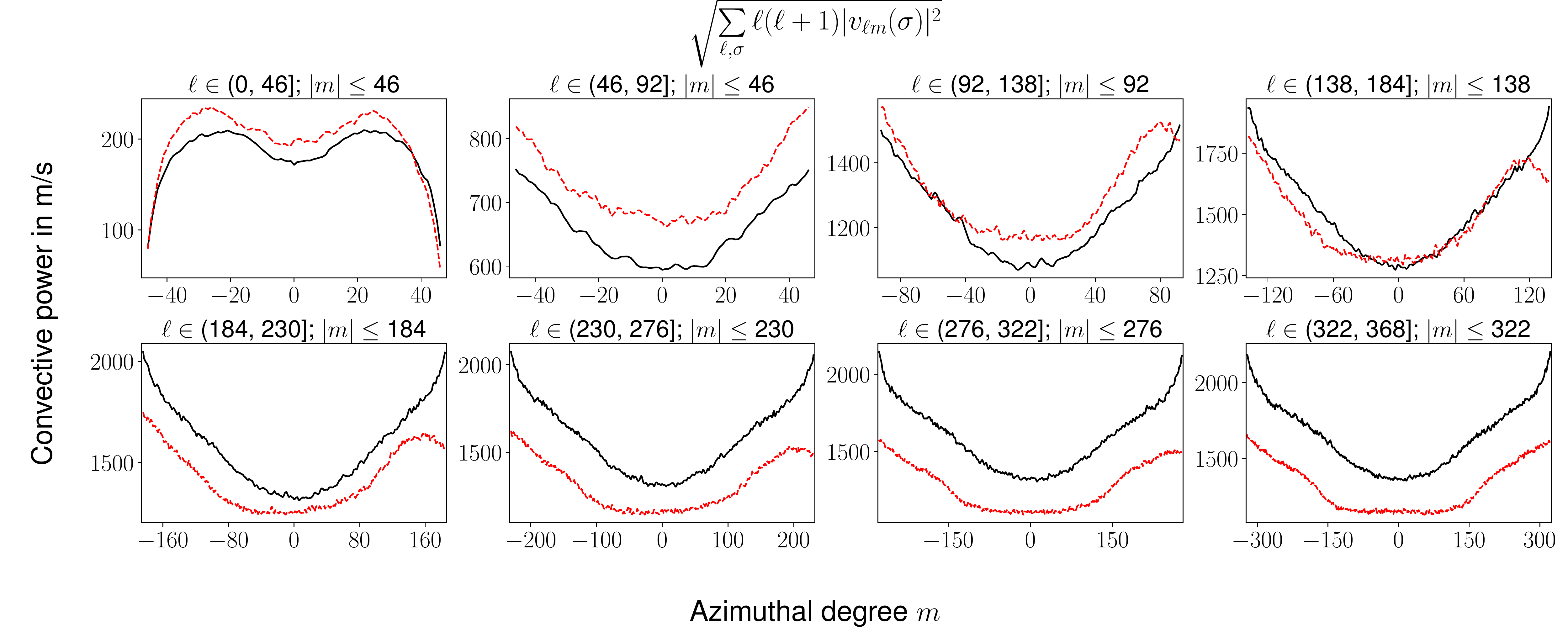}
 \caption{Distribution of poloidal flow power 
 as a function of azimuthal order $m$, inferred 
 from LCT (dashed, red) and the current work (solid, black). LCT has a 
 pronounced plateau at low $m$. The inversions are in good 
 agreement to LCT up to supergranular wavenumbers.}
 \label{fig:figv-t}
\end{figure}

\begin{figure}[!ht]
 \centering
 \includegraphics[width=\textwidth]{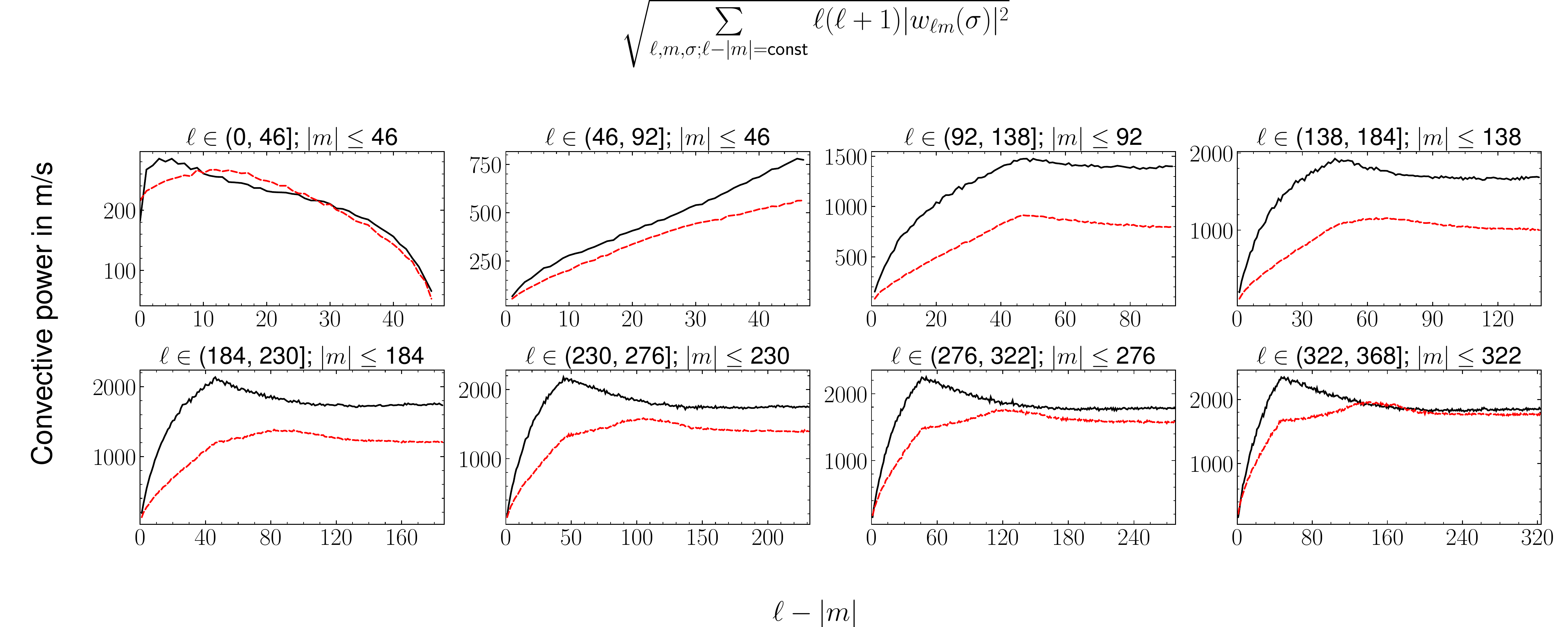}
 \caption{Distribution of toroidal flow power as inferred 
from LCT (dashed, red) and the current work (solid, black). The 
inversions underestimate the power and agree qualitatively with 
LCT up to supergranular wavenumbers. For larger wavenumbers, 
the LCT spectrum is flat at low $m$.}
\label{fig:figw-s-|t|}
\end{figure}

\begin{figure}[!ht]
 \centering
 \includegraphics[width=\textwidth]{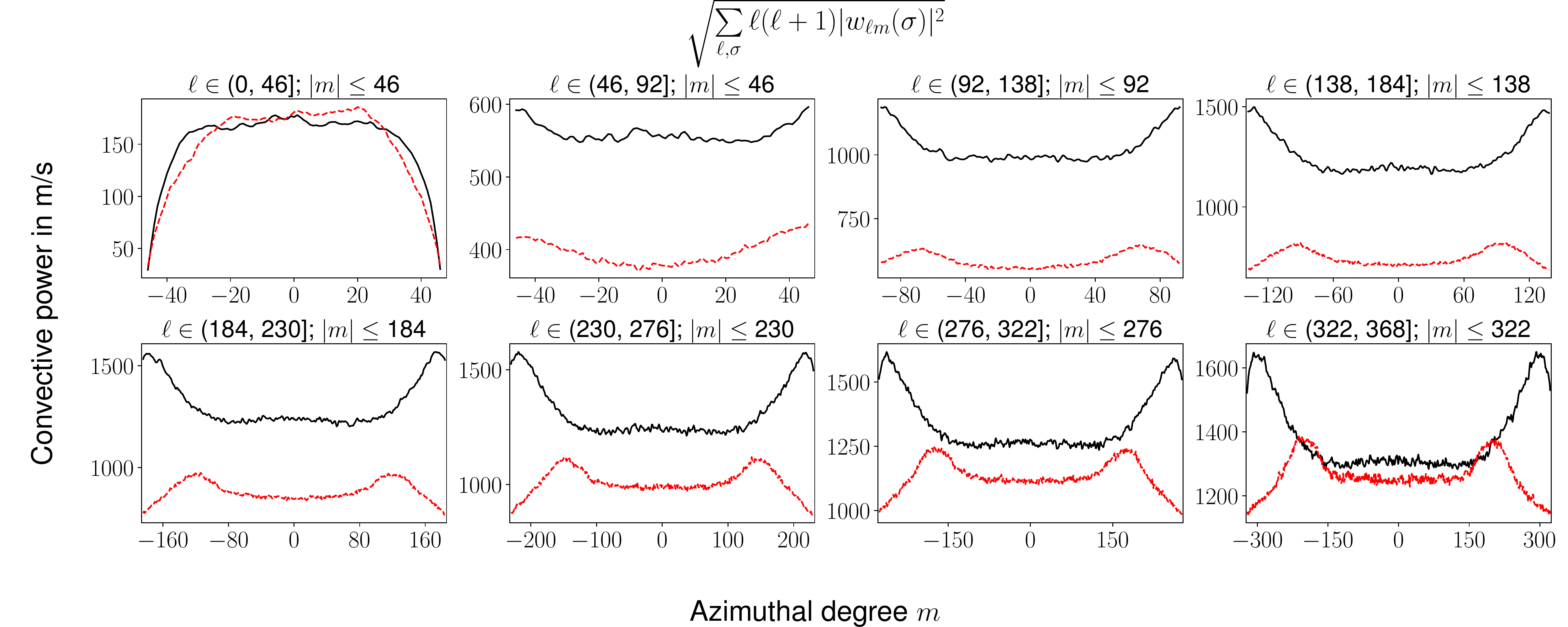}
 \caption{Distribution of toroidal flow power 
 as a function of azimuthal order $m$, inferred 
 from LCT (black) and the current work (red). The LCT power
 peaks at intermediate $m$ whereas inversions suggest a preference
 towards extreme $m$.}
 \label{fig:figw-t}
\end{figure}

As demonstrated in Appendix~\ref{appendix:synth-test}, the synthetic tests
show good poloidal flow reproduction. This can also be seen in the 
comparison with LCT data as shown in Figs.~(\ref{fig:figv-s-|t|}, 
\ref{fig:figv-t}). The toroidal flow inversions are qualitatively 
different from LCT data as seen in Figs.~(\ref{fig:figw-s-|t|},
\ref{fig:figw-t}). As the synthetic tests also suggest, toroidal flow
inversions are the least trustworthy. 

%\begin{figure}[!ht]
% \centering
% \includegraphics[width=\textwidth]{figv_sigma.pdf}
% \caption{Toroidal convective power as function of temporal %frequency.
% LCT (black) shows a dominant peak near $1$ $\mu$Hz for %supergranular
% length scales, but the present work (red) has no dependence on %temporal frequency.}
% \label{fig:figv-sigma}
%\end{figure}

%\begin{figure}[!ht]
% \centering
% \includegraphics[width=\textwidth]{figw_sigma.pdf}
% \caption{Toroidal convective power as function of temporal %frequency.
% LCT (dashed, red) shows a slightly less pronounced peak near $1$ %$\mu$Hz,
% when compared to poloidal flow inversions (solid, black). The %present work
% has no dependence on temporal frequency.}
% \label{fig:figw-sigma}
%\end{figure}

% figv s-|t|

% figv-t

% figw s-|t|

% figw-t

\section{Contamination of radial flows by poloidal flows}
\label{sec:contamination}
\begin{figure}[!ht]
 \centering
 \includegraphics[width=0.6\textwidth]{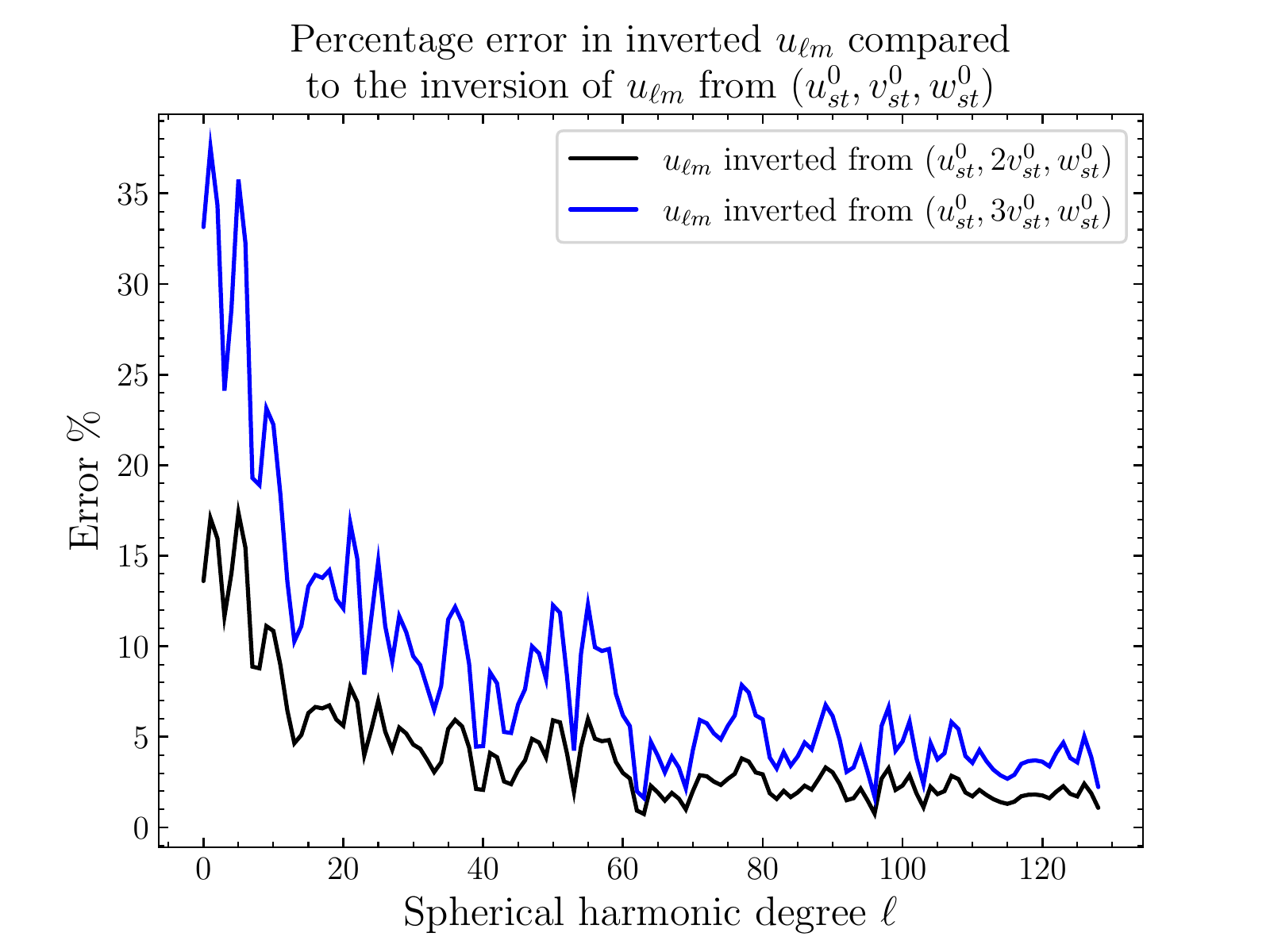}
 \caption{Error in the inference of radial flow when the synthetic poloidal
 flows are increased by 100\% and 200\%. It can be seen that the error reduces
 for higher $\ell$ and is $<10$\% for $\ell > 20$. The error is also 
 $<5\%$ near the supergranular scales.}
 \label{fig:contamination}
\end{figure}

To test the contamination of radial flows by poloidal flows, 
we setup a synthetic test where we perform inversions for
3 different profiles. 
\begin{enumerate}
    \item $(u^0_{st}, v^0_{st}, w^0_{st})$ - maximum $v^0_{st} = 240$ ms${}^{-1}$, 
    maximum $u^0_{st} = 10$ ms${}^{-1}$.
    \item $(u^0_{st}, 2v^0_{st}, w^0_{st})$ - maximum $v^0_{st} = 480$ ms${}^{-1}$, 
    maximum $u^0_{st} = 10$ ms${}^{-1}$.
    \item $(u^0_{st}, 3v^0_{st}, w^0_{st})$ - maximum $v^0_{st} = 720$ ms${}^{-1}$, 
    maximum $u^0_{st} = 10$ ms${}^{-1}$.
\end{enumerate}
We plot the error in the inferred $u_{\ell m}$ for cases 2 and 3 relative to 
the $u_{\ell m}$ of case 1, in Fig~\ref{fig:contamination}. 
For most $\ell$, the error in the estimation of 
the radial flow is $<10\%$, in spite of the poloidal flow having changed by
$200\%$. This shows that the contamination of radial flows by poloidal flows 
is minimal.

\newpage

% bibliography
\bibliography{main}{}
\bibliographystyle{aasjournal}
\end{document}